\documentclass[prl,twocolumn,groupeaddress]{revtex4-2}

\usepackage{amsmath,amssymb}
\usepackage{physics}
\usepackage{graphicx}
\usepackage{physics}
\usepackage{bbold}
\usepackage[colorlinks=true, allcolors=blue]{hyperref}

\begin{document}
\title{Theory of magnetic excitations in multilayer nickelate superconductor La\textsubscript{3}Ni\textsubscript{2}O\textsubscript{7}}

\author{Steffen B\"otzel}
\affiliation{Theoretische Physik III, Ruhr-Universit\"at Bochum,
  D-44780 Bochum, Germany}

\author{Frank Lechermann}
\affiliation{Theoretische Physik III, Ruhr-Universit\"at Bochum,
  D-44780 Bochum, Germany}  

\author{Jannik Gondolf}  
 \affiliation{Theoretische Physik III, Ruhr-Universit\"at Bochum,
  D-44780 Bochum, Germany} 

\author{Ilya M. Eremin}
\affiliation{Theoretische Physik III, Ruhr-Universit\"at Bochum,
  D-44780 Bochum, Germany}

\begin{abstract}
Motivated by the recent reports of high-$T_c$ superconductivity in La$_3$Ni$_2$O$_7$ under pressure, we analyzed theoretically the magnetic excitations in the normal and the superconducting state in this compound, which can be measured by inelastic neutron scattering or RIXS. We show that the bilayer structure of the spin response allows to elucidate the role of the interlayer interaction and the nature of the Cooper-pairing in a very efficient way. In particular, we demonstrate the key difference between the potential $s_\pm$  and  $d$-wave  gaps, proposed recently, by comparing the corresponding response in the even and odd channels of the spin susceptibility. We show that the mostly interlayer driven bonding-antibonding $s_\pm$ Cooper-pairing produces a single large spin resonance peak in the odd channel only near the $X$ point whereas spin resonances in both the odd and the even channel are predicted for the $d$-wave scenario.
\end{abstract}

\maketitle
%\linenumbers  %adds lines to reference changes

\textit{Introduction.---}
The discovery of unconventional superconductivity in hole-doped thin films of infinite-layer and reduced multilayer nickelates~\cite{Li2019,Pan2021,Osada2020} has stimulated further interest in studying exotic quantum states and potential superconductivity in the so-called Ruddlesden-Popper (RP) series of the nickel-based oxides, denoted as R$_{n+1}$Ni$_n$O$_{3n+1}$, where R refers to rare earth element
and $n$ is the number of consecutive layers.
The most recent breakthrough in this direction are
reports of high-pressure superconductivity around 80 K in 
La$_3$Ni$_2$O$_7$~\cite{sun23,JunHou:117302,zhang2023high,zhou2023evidence,zhang2023effects,wang2023structure,wang2023pressure,dong2023visualization} and around 20 K in La$_4$Ni$_3$O$_{10}$~\cite{sakakibara2023theoretical,li2023signature,zhang2023superconductivity,zhu2024superconductivity}. These exciting discoveries motivated massive theoretical investigation~\cite{lechermann2023electronic,liu2023role,oh2023type,luo2023high,qin2023high,huang2023impurity,qu2023bilayer,zhang2023trends,yang2023possible,zhang2023structural,yang2023strong,heier2023competing,jiang2023high,ryee2023critical,tian2023correlation,liao2023electron,kaneko2023pair,LuoModel,chen2023orbital,shen2023effective,yang2023minimal,wu2023charge,lu2023interplay,shilenko2023correlated,zhang2023electronic,geisler2023structural,lange2023feshbach,labollita2023electronic,rhodes2023structural,zhang2023strong,qu2023bilayer,lu2023interlayer,chen2023critical,cao2023flat,christiansson2023correlated,zheng2023superconductivity,kakoi2023pair,fan2023superconductivity,geisler2024optical,sakakibara2024possible}, yet the full structural characterization of these systems is still under debate~\cite{puphal2023unconventional,chen2023polymorphism,zhou2023evidence}. Nevertheless, a striking increase of the superconducting transition temperature in La$_3$Ni$_2$O$_7$ with multi- or bilayer structure calls for a careful theoretical examination.

Considering La$_3$Ni$_2$O$_7$ as RP bilayer systems yields a formal Ni $3d^{7.5}$ (or $3d^8$ when considering ligand-hole physics~\cite{lechermann2023electronic}) electronic configuration with both Ni-$e_{g}$ orbitals crossing the Fermi level. The low energy physics in this system is ruled by the multiorbital and the bilayer effects with strong hybridization between the Ni-$d_{z^2}$ and the apical O-$p_z$ orbitals \cite{zhang2023electronic}. The multiorbital structure seems to be one of the key differences between La$_3$Ni$_2$O$_7$ and the bilayer cuprate superconductors where Cu$^{2+}$ ions with $3d^9$ configuration possess only one unpaired valence electron in the $3d_{x^2-y^2}$-orbital, whereas the Ni ion has unpaired 
valence electrons in both the $3d_{x^2-y^2}$ and $3d_{z^2}$ orbitals.
Various Hubbard-Hund-type or $t-J$ like models have already been proposed to capture the superconducting and normal state properties of this multiorbital system~\cite{lechermann2023electronic,liu2023role,qin2023high,luo2023high,huang2023impurity,oh2023type,qu2023bilayer,ryee2023critical,tian2023correlation,liao2023electron,kaneko2023pair,LuoModel,chen2023orbital,jiang2023high,shen2023effective,yang2023minimal,wu2023charge}. 

Within the variety of model considerations, one of the most interesting theoretical question concerns the interplay between the intralayer and the interlayer Cooper pairing \cite{dagotto1992superconductivity}, which yields a competition between the $s_{\pm}$-wave symmetry of the superconducting order parameter, driven mostly by the interlayer Cooper-pairing~\cite{liu2023role,lechermann2023electronic,qin2023high,zhang2023trends,yang2023possible,luo2023high,oh2023type,qu2023bilayer,zhang2023structural,huang2023impurity,heier2023competing,yang2023strong,sakakibara2024possible} and the $d_{x^2-y^2}$-wave or the $d_{xy}$-wave symmetries of the superconducting order parameters, driven mostly by the intralayer interaction, respectively~\cite{lechermann2023electronic,liu2023role,jiang2023high,heier2023competing,lu2023interlayer}. 

Given the likely non-phononic origin of superconductivity in pressurized La$_3$Ni$_2$O$_7$, the strange metal behavior of the normal state~\cite{zhang2023high,sun23,wang2023pressure} and the signatures of magnetic ordering, seen at the ambient pressure~\cite{chen2023evidence,liu2023evidence} at around 150 K, it is instructive to study the bilayer-structure impact on the spin response in this system in the normal and superconducting states. Recall that one of the important hallmarks of the superconducting state in unconventional superconductors is the occurrence of the so-called spin resonance. It is seen by the inelastic neutron scattering in various systems at the antiferromagnetic (AFM) wave vector $\mathbf{Q}_{\rm AFM}= (\pi,\pi)$ at energies below or around the superconducting gap energy threshold of about 2$\Delta$~\cite{yu2009universal}. Its presence in various unconventional superconductors ranging from high-T$_c$ cuprates~\cite{rossat1991neutron,mookPolarized,Bourges1996,Dai1999,Hinkov2007}, iron-based superconductors~\cite{Christianson2008,Inosov2009,Park2011,Dai2015}, and some heavy-fermion superconductors like CeCoIn$_5$~\cite{Stock2008,Song2020} is considered to be one of the strong signatures of spin-fluctuation mediated Cooper-pairing~\cite{yu2009universal,scalapino2012common}. In the simplest theoretical picture, the spin resonance peak occurs due to a change of sign of the superconducting order parameter at the parts of the Fermi surface which are connected by the AFM wave vector $\Delta_\mathbf{k}$ and $\Delta_{\mathbf{k}+\mathbf{Q}}$~\cite{bulut1992nodeless,Fong1996,Abanov1999,Brinkmann1999,Norman2000,Kao2000,Chubukov2001,Onufrieva2002,Eremin2005}. 

The occurrence of the resonance peaks and their dispersion in bilayer systems not only allows to confirm the unconventional nature of superconductivity and to learn about the superconducting gap symmetry, but also to understand the importance of the interlayer coupling. In particular, due to two CuO$_2$ layers per unit cell, the spin response in bilayer cuprates splits into even and odd channels. This in turn can be connected to the bonding and anti-bonding character of the electronic bands, showing modulations in $q_z$ directions~\cite{sato1988two}. The magnetic susceptibility splits into even $\chi_{e}$ and odd $\chi_o$ susceptibilities~\cite{blumberg1995antiferromagnetic,Normand,eremin2007spin,headings2011spin} and they can be accessed individually by measuring the spin response at different $q_z$ momenta. In bilayer cuprates, the spin resonance peak was found in both channels in a wide doping range with clear splitting between $\chi_e$ and $\chi_o$~\cite{Pailhes2006,Capogna2007}. Furthermore, analyzing their intensities and frequency positions, one was able to extract important information of the overall structure of the paramagnon continuum and the interplay of interlayer exchange interactions and interlayer hopping matrix elements~\cite{blumberg1995antiferromagnetic,Normand,eremin2007spin,headings2011spin}. In La$_3$Ni$_2$O$_7$ the situation is even more intriguing, because the splitting of bonding and antibonding bands is much stronger.

In this work, we provide a theoretical description of the spin response of paramagnetic La$_3$Ni$_2$O$_7$ in the pressurized normal and superconducting state. We begin by considering the spin response in multiorbital bilayer systems and then focus on two different scenarios for the superconducting pairing. First, we consider $s_\pm$ interlayer pairing with a sign change between bonding and anti-bonding bands. Using the tight-binding model from Ref.~\cite{LuoModel} we discuss the important scattering vectors contributing to $\chi_e$ and $\chi_o$. We find that $\chi_o$ gives a much larger spin response. In the superconducting state, this scenario manifests in a dominant spin resonance peak near the X-point. The second scenario is a cuprate like $d_{x^2-y^2}$ gap symmetry which is discussed in the context of the model presented in Ref.~\cite{lechermann2023electronic}. Here, the spin response for $\chi_e$ and $\chi_o$ channels is of the same order. Correspondingly, this scenario yields different spin resonance peaks in the superconducting state in both $\chi_e$ and $\chi_o$. Measuring the spin response in both channels can therefore clearly reveal the actual Cooper-pairing scenario and clarify the role of the interlayer exchange interaction in this system.

\begin{figure*}[t]
      \includegraphics[width=\linewidth]{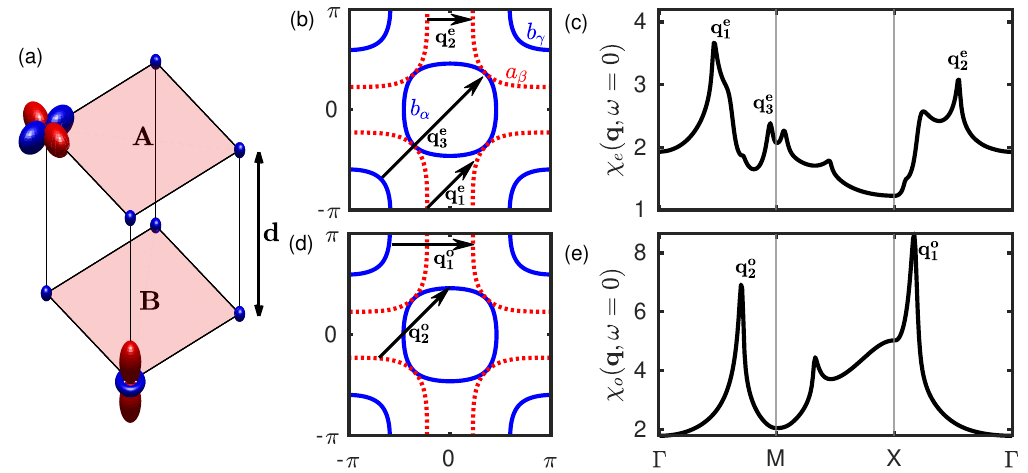}
	\caption{
  (a) Schematic representation of the bilayer model with $e_g$ orbitals.
  (b) and (d) show the Fermi surface for the model~\cite{LuoModel} with bonding $(b)$ and anti-bonding $(a)$ bands shown by blue (solid) and red (dashed) curves, respectively. Note the sign-changing bonding-antibonding $s_\pm$-gap follows the same red-blue color distribution referring to the negative and positive regions of the gap. Greek indices denote the notion of the bands, accepted in literature. The arrows in (b) and (d) highlight the important scattering wavevectors in the even and odd channels, respectively. (c) and (e) show the even ($\chi_e$) and the odd ($\chi_o$) parts of the static RPA susceptibility in arbitrary units, respectively, calculated in the normal state ($T = 80$ K) for $U = 0.9 U_{\text{mag}}$ and $J_{\text{H}}=U/7$. Peaks are labeled according to scattering vectors displayed in (b) and (d).} \label{fig1}
\end{figure*}

\textit{Theoretical Approach.---}
We consider an effective quasi-two-dimensional bilayer Hamiltonian for the Ni-$e_g$ orbitals, illustrated in Fig.~\ref{fig1}(a), which follows from the tight-binding models developed previously~\cite{LuoModel,lechermann2023electronic}. \textcolor{black}{The considered Hamiltonian consists of non-interacting and multiorbital on-site interacting parts $\mathcal{H} = \mathcal{H}_0 + \mathcal{H}_{\text{int}}$. Note that the combination of the large interlayer hopping of the system and the on-site interaction results in an effective interlayer interaction corresponding to the superexchange process through the inner apical oxygen.}  
%%%%%%%%%%%%%%%%%%%%%%%%%%%%%%%%%%%%%%%%%%%%%%%%%%%%%%%%%%%%%
%%%%%%%%%%%%%%%%%%%%%%%%%%%%%%%%%%%%%%%%%%%%%%%%%%%%%%%%%%%%%%
\textcolor{black}{In momentum space the non-interacting part reads}
\begin{equation}
   \mathcal{H}_{0} = \sum_{\mathbf{k},k_z}\sum_{l_1,l_2}\sum_{o_1,o_2} \hat{H}_{l_1o_1;l_2o_2}(\mathbf{k},k_z) c^{\dagger}_{l_1o_1}(\mathbf{k},k_z)c_{l_2o_2}(\mathbf{k},k_z),  
\end{equation}
where $c^{\dagger}_{l,o}(\mathbf{k},k_z)$ creates an electron in layer $l$ and orbital $o$ with in-plane momentum $\mathbf{k}=(k_x,k_y)$. \textcolor{black}{In our modeling the out-of-plane momentum $k_z$ only enters in the phase factors arising from the Fourier transform for a bilayer system. We denote the upper and the lower layer with $A$ and $B$, respectively. As the Hamiltonian must be invariant under exchange of layers, we find $\hat{H}_{AA} = \hat{H}_{BB} =  \hat{H}_\parallel$  and $\hat{H}_{AB} = \hat{H}_{BA} = \hat{H}_{\perp}$ for the intralayer and interlayer hoppings, respectively, yielding the general form}
\begin{equation}
\hat{H}(\mathbf{k},k_z) =
    \begin{pmatrix}
		\hat{H}_{\parallel}(\mathbf{k})	
        & \hat{H}_{\perp}(\mathbf{k}) e^{ik_zd} \\
		\hat{H}_{\perp}(\mathbf{k}) e^{-ik_zd}
        & \hat{H}_{\parallel}(\mathbf{k})	
	\end{pmatrix}.
 \label{eq:bilayerHamiltonian}
\end{equation}
Here the $k_z$ dependence only appears due to the phase factors with $d$ being the height of the bilayer sandwich. The hats are put to remind the reader that all blocks are, in principle, square matrices in the orbital degree of freedom. The above Hamiltonian can be block-diagonalized using the transformation
\begin{align}
V = \frac{1}{\sqrt{2}}
    \begin{pmatrix}
		\mathbb{1}	
        &   \mathbb{1} e^{ik_zd} \\
		 \mathbb{1} e^{-ik_zd}
        & -\mathbb{1}
	\end{pmatrix},
 \ \ \ \
 \hat{H}^{b/a} =  \hat{H}_{\parallel} \pm \hat{H}_{\perp},
 \label{eq:abTrafo}
\end{align}
where the blocks belong to the bonding (b) and anti-bonding (a) sub-spaces. Note that the phase factor is not present in the (ab)-space.

For the interaction part of the Hamiltonian we include the on-site  intraorbital ($U$) and interorbital ($U'$), Hund's type ($J_{\text{H}}$) and pair hopping ($J'$) interaction and assume spin-rotational
invariance, which yields the relations $U' = U - 2J_{\text{H}}$ and $J_{\text{H}} = J'$ \cite{lechermann2023electronic,graser2009near}. 

The non-interacting multiorbital susceptibility in the superconducting state can be written in terms of normal and anomalous Green's functions
\begin{align}
    (\chi_0)_{\eta_1\eta_4}^{\eta_2\eta_3}(q) &= \frac{T}{N} \sum_{k} 
[ F_{\eta_1\eta_3}(k+q) \bar{F}_{\eta_2\eta_4}(k) \nonumber \\ &- G_{\eta_1\eta_2}(k+q) G_{\eta_3\eta_4}(k) ],
\end{align}
where we use the four notation $k=(\mathbf{k},k_z,i\omega_n)$ and shorthand indices $\eta = (l,o,s)$. To compute the spin susceptibility, we use the random phase approximation (RPA)~\cite{lechermann2023electronic,graser2009near}. The interacting susceptibility can be written as a matrix equation with susceptibility matrices of the form
\begin{align}
    \hat{\chi}(q_z) &=
    \begin{pmatrix}
		\hat{\chi}_{\parallel}	
        & \hat{\chi}_{\perp}	e^{iq_zd}  \\
		\hat{\chi}_{\perp} e^{-iq_zd}
        & \hat{\chi}_{\parallel}
    \end{pmatrix},
    \label{eq:diagonalBilayerSubspaceSusceptibility} 
\end{align}
\textcolor{black}{where we suppressed the in-plane momentum dependence $\mathbf{q}$ and the dependence on Matsubara frequencies and the $q_z$ dependence only enters via the phase factors.} 
The Dyson-type RPA matrix equation can be decomposed into even and odd channels with respect to the exchange of layer index by defining $\hat{\chi}^{e/o} = 2( \hat{\chi}_{\parallel} \pm \hat{\chi}_{\perp})$:
\begin{equation}
	\label{Eq:RPAMatrixEqEvenOdd}
	\hat{\chi}^{(e/o)}
	=
	\left[
	\mathbb{1} -
    \frac{1}{2}
	\hat{\chi}^{(e/o)}_{0}
	\hat{U}
	\right]^{-1}
    \hat{\chi}^{(e/o)}_{0},
\end{equation}
where $\chi_0$ denotes the non-interacting susceptibilities. The above expression holds in general for both, the spin and the charge susceptibilities, but the interaction matrix $\hat{U}$ has to be chosen differently. \textcolor{black}{For the physical paramagnetic susceptibility we have to contract the orbital and sublattice degrees of freedom at the free vertices. In terms of $\hat{\chi}_e$ and $\hat{\chi}_o$ it can be written as} 
\begin{align}
    \chi_{\text{spin}}
    &=
    \sum_{l_1l_2,o_1o_2}
    \begin{pmatrix}
		\hat{\chi}^{e}+\hat{\chi}^o & (\hat{\chi}^{e}-\chi^o) e^{iq_zd}  \\
		  (\hat{\chi}^{e}-\hat{\chi}^o) e^{-iq_zd} & \hat{\chi}^{e}+\hat{\chi}^o
	\end{pmatrix}_{l_1o_1;l_2o_2} 
 \nonumber \\
   &= \sum_{o_1,o_2} [ \hat{\chi}^{e}_{o_1,o_2} \cos^2(\frac{q_z d}{2}) + \hat{\chi}^{o}_{o_1,o_2} \sin^2(\frac{q_z d}{2}) ].
   \label{eq:SpinSusceptibility}
\end{align}
This expression of the spin susceptibility has been initially derived for the bilayer cuprates \cite{sato1988two,Brinkmann1999,eremin2007spin}.  
By explicitly using the matrix elements from Eq.~\ref{eq:abTrafo}, we can express the even and the odd susceptibilities through susceptibilities in the $(ab)$-space:
$\hat{\chi}^{e}_0 = \hat{\chi}^{aa} + \hat{\chi}^{bb}$ and $\hat{\chi}^{o}_0 = \hat{\chi}^{ab} + \hat{\chi}^{ba}$. \textcolor{black}{More information on the theoretical approach and used parameters can be found in the supplementary.}

%%%%%%%%%%%%%%%%%%%%%%%%%%%%%%%%%%%%%%%%%%%%%%%%%%%%%%%%%%%%%%%%%%
\textit{Results.---} In what follows we compute the bilayer spin response for pressurized La$_3$Ni$_2$O$_7$ for two slightly different tight-binding parameterizations of the non-interacting Hamiltonian to observe the general trends. In Fig.~\ref{fig1}(c),(e) and Fig.~\ref{fig3}(c) we show the normal state even and odd components of the bilayer spin susceptibility using the tight-binding model from Ref.~\cite{LuoModel} and Ref.~\cite{lechermann2023electronic}, respectively. One immediately sees that independent of the model used, the magnetic response in the odd and the even channels strongly differ from each other with respect to the dominant scattering peaks. \textcolor{black}{Specifically, the main scatterings in the even channel stem from the scattering within antibonding $\beta$ band and within bonding $\gamma$ band ({\bf q}$_1^e$ and {\bf q}$_2^e$) as well as scattering between bonding $\alpha$ and $\gamma$ bands ({\bf q}$_3^e$), which is illustrated in Fig.~\ref{fig1}(b). At the same time, the main scatterings in the odd channel stem from scattering between bonding $\alpha$ to antibonding $\beta$ band ({\bf q}$_2^o$) and bonding $\gamma$ band to antibonding $\beta$ band ({\bf q}$_1^o$)}, which is displayed in Fig.~\ref{fig1}(d). Apart from different magnitudes, these peaks are dominant in both models. \textcolor{black}{The different behavior in both channels arises from the large bonding-antibonding splitting of the electronic bands as can be seen by looking on the non-interacting susceptibility presented in Fig.1 of the supplementary and does not require the effective interlayer interaction.}

Despite the similar {\bf q}-dependence of the spin response in both models, the dominant superconducting instabilities appear to be different. While models based on Ref.~\cite{LuoModel} and spin-fluctuation mediated pairing mostly predict sign-changing bonding-antibonding $s$-wave solutions~\cite{oh2023type,luo2023high}, similar spin-fluctuation based analyses reveal $d$-wave symmetries to be present in the other model~\cite{lechermann2023electronic,liu2023role}. One can understand this difference by simply looking into the overall magnitudes of the even and odd susceptibilities. In the case of the tight-binding model of Ref.~\cite{LuoModel} the odd susceptibility appears to be twice larger in magnitude than the even susceptibility, which indicates the dominance of the interlayer itinerant magnetic fluctuations. This dominance of the magnetic fluctuations in the odd channel supports the interlayer (bonding-antibonding) $s_\pm$-wave Cooper-pairing with large interlayer superconducting gap~\cite{lu2023interplay,yang2023minimal}. At the same time, within the other tight-binding model~\cite{lechermann2023electronic} the different peaks in the odd and the even susceptibilities appear to have similar magnitudes and among several candidates, the $d$-wave superconducting gap appears to be the most stable solution. One should further note at this point that the $d$-wave symmetry is usually more stable against the inclusion of the local Coulomb interaction and even in the case of near competition between various channels this solution usually wins~\cite{coleman2019coulomb}. 

The strong differentiation of the even and the odd susceptibilities seen in the normal state offer now a practical tool to disentangle between various Cooper-pairing scenarios. For this, we now extend the calculations of the spin response to the superconducting state by comparing the structure of the spin response for the model giving the sign-changing bonding-antibonding $s$-wave solution~\cite{LuoModel} and that, which gives the $d$-wave solution to be the most dominant~\cite{lechermann2023electronic}. 
 
Note, the bonding-antibonding $s_{\pm}$-wave solution in the (ab)-space allows for a simple decomposition of the Cooper-pairing to the interlayer and intralayer contribution. In particular, we can write the gaps in the sublattice space as $\hat{\Delta}^{a/b} = \hat{\Delta} \pm \hat{\Delta}^\perp$, where $\hat{\Delta}^\parallel$ and $\hat{\Delta}^\perp$ are the intralayer and the interlayer superconducting order parameters, respectively. By setting $\Delta_{x^2-y^2}^\perp=\Delta_{z^2}^\perp$ and $\hat{\Delta}^\parallel=0$, one ends up with a constant magnitude gap, which however changes sign between bonding and anti-bonding bands. This gap symmetry is consistent with the numerical solutions found in the literature~\cite{liu2023role,lechermann2023electronic,qin2023high,zhang2023trends,yang2023possible,luo2023high,oh2023type,qu2023bilayer,zhang2023structural,huang2023impurity,heier2023competing,yang2023strong} and is illustrated in Fig.~\ref{fig1}(b). 

\begin{figure}[t]
      \includegraphics[width=\linewidth]{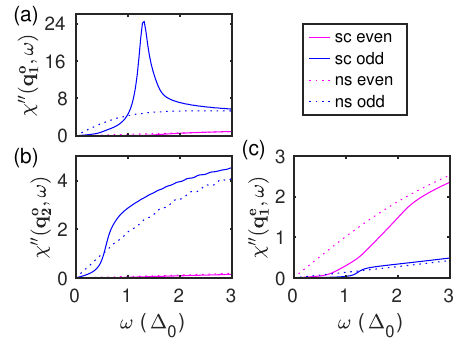}
	\caption{
Calculated frequency dependence of the imaginary part of the even and the odd spin susceptibility in the bonding-antibonding s$_{\pm}$-wave superconducting (solid curves) and normal (dashed curves) states using the tight-binding parameters from Ref.~\cite{LuoModel}. The representative wave vectors $\mathbf{q}_1^{o}$ (a), $\mathbf{q}_2^{o}$ (b) and $\mathbf{q}_1^{e}$ (c) are chosen from Fig.~\ref{fig1}.
} \label{fig2}
\end{figure}

\begin{figure}[t]
      \includegraphics[width=\linewidth]{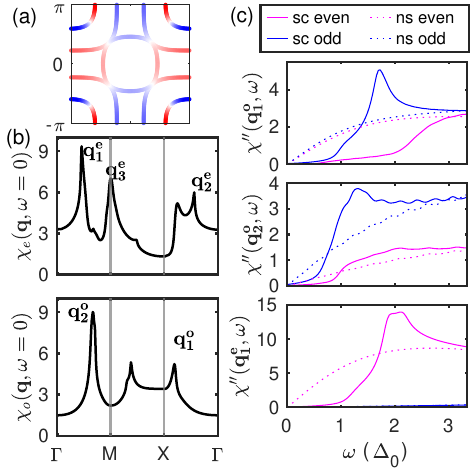}
	\caption{
  (a) Sketch of the $d_{x^2-y^2}$-wave gap solution for the tight-binding model~\cite{lechermann2023electronic} where blue and red colors refer to the opposite signs of the gap magnitudes. (b) shows the even ($\chi_e$) and the odd ($\chi_o$) parts of the static RPA susceptibility, respectively, calculated in the normal state ($T = 80$ K) for $U = 0.9 U_{\text{mag}}$ and $J_{\text{H}}=U/4$. (c) shows the calculated frequency dependence of the imaginary part at the characteristic maximal scattering wavevectors of the even and the odd spin susceptibility for the $d_{x^2-y^2}$-wave superconducting (solid curves) and normal (dashed curves) states and the maximum gap magnitude of $\Delta_0 =$15~meV. \textcolor{black}{Note that at the bottom panel the blue lines are near 0.}
} \label{fig3}
\end{figure}

Given the peculiar structure of the sign-changing bonding-antibonding gap, the spin resonance in the spin susceptibility exclusively appears in the odd channel with the scattering between bonding and antibonding bands but not in the even channel. In particular, we show in Fig.~\ref{fig2} the calculated frequency dependence of the imaginary parts of the even, Im$\chi_{e}$,  and the odd, Im$\chi_{o}$, spin susceptibilities at the characteristic wave vectors $\mathbf{q}_1^{o}$, $\mathbf{q}_2^{o}$, and $\mathbf{q}_1^{e}$ identified in the normal state and employing the characteristic gap size of $\Delta_0 = 15 $~meV, which yields a plausible gap-to-$T_c$ ratio of about 4.3. Observe that the enhancement occurs only for the odd spin response at $\mathbf{q}_1^{o}$ and $\mathbf{q}_2^{o}$  where the superconducting gap changes sign between bonding and antibonding bands and the exact magnitude depends on the values of the superconducting gap at the corresponding region. On the contrary, the spin susceptibility in the even channel at $\mathbf{q}_1^{e}$ is suppressed as generally the scattering wavevectors within bonding or antibonding bands connect regions of the same gap sign of the superconducting order parameter. In this regard, the outlined behavior of the spin response, i.e. the enhancement of the odd spin susceptibility and its absence in the even channel in the experiment probed by inelastic neutron scattering or RIXS will be a direct probe for the interlayer Cooper-pairing and bonding-antibonding character of the $s_{\pm}$-wave gap. 

\textcolor{black}{The obtained results are robust to small variations of the gap. The solutions discussed in the literature sometimes show nodal regions appearing in $\Gamma-M$ direction, which can be incorporated by fine-tuning and inclusion of interorbital gaps or by including the gaps directly in band space, which we have done for the results shown. However, we find that such details are essentially irrelevant for the generic behavior of the spin resonance. Three different $s_\pm$ gap functions are compared exemplarily in the supplementary.}
%%%%%%%%%%%%%%%%%%%%%%%%%%%%%%%%%%%%%%%%%%%%%%%%%%%%%%%%%%%%%
%%%%%%%%%%%%%%%%%%%%%%%%%%%%%%%%%%%%%%%%%%%%%%%%%%%%%%%%%%%%%%

%Let us now turn to the second scenario of intralayer pairing for which we assume a $d_{x^2-y^2}$ symmetry, i.e. we use $\Delta^{x/z} = \pm \Delta_0 \left[ \cos(k_x)-\cos(k_y) \right]$.
Let us now turn to the discussion of the spin response for the second of the models outlined in Ref.~\cite{lechermann2023electronic}, which give the $d_{x^2-y^2}$-wave superconducting gap symmetry, shown in Fig.~\ref{fig3}(a). Here we should mention that  \textcolor{black}{due to the mixed orbital character of the $\alpha$ and the $\beta$ band the $d_{x^2-y^2}$-wave solution cannot be straightforwardly decomposed in the orbital and sublattice degrees of freedom. Instead we introduce the gap function in the band space and having in mind that it has different contributions including those from interorbital and inter- and intralayer gaps, which is required to avoid interband gaps.} 
%%%%%%%%%%%%%%%%%%%%%%%%%%%%%%%%%%%%%%%%%%%%%%%%%%%%%%%%%%%%%
%%%%%%%%%%%%%%%%%%%%%%%%%%%%%%%%%%%%%%%%%%%%%%%%%%%%%%%%%%%%%%

In contrast to the bonding-antibonding s$_{\pm}$-wave scenario, a sizeable enhancement of  Im$\chi$ in the $d$-wave superconducting state is seen in the odd channel at $\mathbf{q}_1^{o}$ as well as in the even channel near $\mathbf{q}_1^{e}$ as shown in Fig.~\ref{fig3}(c). The differences in the overall magnitude can be traced back to the angular dependence of the superconducting gap at the Fermi surface. For example, there is no sign change of the gap on the $\beta$ Fermi surface sheet portions for the wavevectors ${\bf q}_1^e$ and ${\bf q}_2^e$, shown as arrows in Fig.~\ref{fig1}(b). In fact, it is the scattering across the $\gamma$-pocket that causes the enhancement and happens to be peaked at the same momentum transfer. \textcolor{black}{The large intraband contributions from the $\gamma$ band to the susceptibility can be attributed to its flatness.} In contrast to the odd spin resonance in the s$_{\pm}-$wave scenario, the resonances in the odd and even channels in the $d_{x^2-y^2}$-wave scenario are much broader and for the even channel also strongly dispersing in the momentum space\textcolor{black}{, which is shown in Fig.2(b) of the Supplementary.} Most importantly, within the $d-$wave scenario there is no \textcolor{black}{sizeable difference in the magnitude of the enhancements of the spin response in the superconducting state in the odd and the even channel}. 

Let us also mention that in Refs.~\cite{lechermann2023electronic,liu2023role,heier2023competing} the $d_{xy}$-wave solution was also appearing as one of the leading instability for smaller Hund's coupling. Similar to the $d_{x^2-y^2}$ case we find a sharp spin resonance peak in the odd channel at $\mathbf{q}_1^{o}$ and a broad peak the even channel with the dispersion along $\Gamma-X$ direction\textcolor{black}{, which is shown in Fig.2(c) of the supplementary}, while in the  $d_{x^2-y^2}$ one finds the dispersion along $\Gamma-M$-direction. Therefore $d_{xy}$ and $d_{x^2-y^2}$ pairing symmetries can be distinguished by comparing $\Gamma-X$ and $\Gamma-M$ directions for the even channel of the spin susceptibility.

A different Fermi surface topology in which the $\gamma$ pocket sinks below the Fermi surface has been predicted in Ref.~\cite{ryee2023critical}, which is in line with the recent ARPES results for ambient pressure~\cite{yang2023orbital}. For the bonding-antibonding $s_\pm$-wave Cooper-pairing scenario, which changes sign between $\alpha$ and $\beta$ bands, a spin resonance would appear exclusively at $\mathbf{q}_2^{o}$. 
The scattering vector $\mathbf{q}_2^{o}$ may also play an important role in the magnetic ordering at ambient pressure. In the reciprocal space index corresponding to the pseudo-tetragonal unit cell (H,K,L), this peak is located at {$(0.34,0.34,c/(2d)\approx2.6)$} close to the magnetic peak (0.25,0.25,2.5) found in very recent RIXS measurements at ambient pressure \cite{chen2024electronic}. We believe that the fact that in experiment this peak was found to be non-dispersive but loosing intensity away from L$=2.5$ towards lower L (see Fig. 2c in Ref.~\cite{chen2024electronic}) is an indicator of being present in the odd channel exclusively. 

{\it Conclusion.}-- We have shown that the interlayer bonding-antibonding $s_\pm$ and $d$-wave pairing scenarios yield clearly distinguishable bilayer spin responses in the normal and the superconducting states. The $s_\pm$-wave Cooper pairing gives rise to a strong spin resonance peak in the odd channel of the spin susceptibility, whereas the even response shows no enhancement. In contrast, for the $d$-wave symmetry spin resonance-like weaker peaks appear in both, even and odd, channels. Therefore, studying bilayer structure of the spin response in the La$_3$Ni$_2$O$_7$ may provide a crucial check for the superconducting gap symmetry. 

\section{Acknowledgements} 
The work is supported by the German Research Foundation within the bilateral NSFC-DFG Project ER 463/14-1. 

\bibliography{literatur}

%apsrev4-2.bst 2019-01-14 (MD) hand-edited version of apsrev4-1.bst
%Control: key (0)
%Control: author (8) initials jnrlst
%Control: editor formatted (1) identically to author
%Control: production of article title (0) allowed
%Control: page (0) single
%Control: year (1) truncated
%Control: production of eprint (0) enabled
\begin{thebibliography}{5}%
\makeatletter
\providecommand \@ifxundefined [1]{%
 \@ifx{#1\undefined}
}%
\providecommand \@ifnum [1]{%
 \ifnum #1\expandafter \@firstoftwo
 \else \expandafter \@secondoftwo
 \fi
}%
\providecommand \@ifx [1]{%
 \ifx #1\expandafter \@firstoftwo
 \else \expandafter \@secondoftwo
 \fi
}%
\providecommand \natexlab [1]{#1}%
\providecommand \enquote  [1]{``#1''}%
\providecommand \bibnamefont  [1]{#1}%
\providecommand \bibfnamefont [1]{#1}%
\providecommand \citenamefont [1]{#1}%
\providecommand \href@noop [0]{\@secondoftwo}%
\providecommand \href [0]{\begingroup \@sanitize@url \@href}%
\providecommand \@href[1]{\@@startlink{#1}\@@href}%
\providecommand \@@href[1]{\endgroup#1\@@endlink}%
\providecommand \@sanitize@url [0]{\catcode `\\12\catcode `\$12\catcode `\&12\catcode `\#12\catcode `\^12\catcode `\_12\catcode `\%12\relax}%
\providecommand \@@startlink[1]{}%
\providecommand \@@endlink[0]{}%
\providecommand \url  [0]{\begingroup\@sanitize@url \@url }%
\providecommand \@url [1]{\endgroup\@href {#1}{\urlprefix }}%
\providecommand \urlprefix  [0]{URL }%
\providecommand \Eprint [0]{\href }%
\providecommand \doibase [0]{https://doi.org/}%
\providecommand \selectlanguage [0]{\@gobble}%
\providecommand \bibinfo  [0]{\@secondoftwo}%
\providecommand \bibfield  [0]{\@secondoftwo}%
\providecommand \translation [1]{[#1]}%
\providecommand \BibitemOpen [0]{}%
\providecommand \bibitemStop [0]{}%
\providecommand \bibitemNoStop [0]{.\EOS\space}%
\providecommand \EOS [0]{\spacefactor3000\relax}%
\providecommand \BibitemShut  [1]{\csname bibitem#1\endcsname}%
\let\auto@bib@innerbib\@empty
%</preamble>
\bibitem [{\citenamefont {Graser}\ \emph {et~al.}(2009)\citenamefont {Graser}, \citenamefont {Maier}, \citenamefont {Hirschfeld},\ and\ \citenamefont {Scalapino}}]{graser2009near}%
  \BibitemOpen
  \bibfield  {author} {\bibinfo {author} {\bibfnamefont {S.}~\bibnamefont {Graser}}, \bibinfo {author} {\bibfnamefont {T.}~\bibnamefont {Maier}}, \bibinfo {author} {\bibfnamefont {P.}~\bibnamefont {Hirschfeld}},\ and\ \bibinfo {author} {\bibfnamefont {D.}~\bibnamefont {Scalapino}},\ }\bibfield  {title} {\bibinfo {title} {Near-degeneracy of several pairing channels in multiorbital models for the fe pnictides},\ }\href@noop {} {\bibfield  {journal} {\bibinfo  {journal} {New Journal of Physics}\ }\textbf {\bibinfo {volume} {11}},\ \bibinfo {pages} {025016} (\bibinfo {year} {2009})}\BibitemShut {NoStop}%
\bibitem [{\citenamefont {Luo}\ \emph {et~al.}(2023)\citenamefont {Luo}, \citenamefont {Hu}, \citenamefont {Wang}, \citenamefont {W\'u},\ and\ \citenamefont {Yao}}]{LuoModel}%
  \BibitemOpen
  \bibfield  {author} {\bibinfo {author} {\bibfnamefont {Z.}~\bibnamefont {Luo}}, \bibinfo {author} {\bibfnamefont {X.}~\bibnamefont {Hu}}, \bibinfo {author} {\bibfnamefont {M.}~\bibnamefont {Wang}}, \bibinfo {author} {\bibfnamefont {W.}~\bibnamefont {W\'u}},\ and\ \bibinfo {author} {\bibfnamefont {D.-X.}\ \bibnamefont {Yao}},\ }\bibfield  {title} {\bibinfo {title} {Bilayer two-orbital model of {$\mathrm{L}{\mathrm{a}}_{3}\mathrm{N}{\mathrm{i}}_{2}{\mathrm{O}}_{7}$} under pressure},\ }\href {https://doi.org/10.1103/PhysRevLett.131.126001} {\bibfield  {journal} {\bibinfo  {journal} {Phys. Rev. Lett.}\ }\textbf {\bibinfo {volume} {131}},\ \bibinfo {pages} {126001} (\bibinfo {year} {2023})}\BibitemShut {NoStop}%
\bibitem [{\citenamefont {Lechermann}\ \emph {et~al.}(2023)\citenamefont {Lechermann}, \citenamefont {Gondolf}, \citenamefont {B\"otzel},\ and\ \citenamefont {Eremin}}]{lechermann2023electronic}%
  \BibitemOpen
  \bibfield  {author} {\bibinfo {author} {\bibfnamefont {F.}~\bibnamefont {Lechermann}}, \bibinfo {author} {\bibfnamefont {J.}~\bibnamefont {Gondolf}}, \bibinfo {author} {\bibfnamefont {S.}~\bibnamefont {B\"otzel}},\ and\ \bibinfo {author} {\bibfnamefont {I.~M.}\ \bibnamefont {Eremin}},\ }\bibfield  {title} {\bibinfo {title} {Electronic correlations and superconducting instability in {${\mathrm{La}}_{3}{\mathrm{Ni}}_{2}{\mathrm{O}}_{7}$} under high pressure},\ }\href {https://doi.org/10.1103/PhysRevB.108.L201121} {\bibfield  {journal} {\bibinfo  {journal} {Phys. Rev. B}\ }\textbf {\bibinfo {volume} {108}},\ \bibinfo {pages} {L201121} (\bibinfo {year} {2023})}\BibitemShut {NoStop}%
\bibitem [{\citenamefont {Setyawan}\ and\ \citenamefont {Curtarolo}(2010)}]{Setyawan_2010}%
  \BibitemOpen
  \bibfield  {author} {\bibinfo {author} {\bibfnamefont {W.}~\bibnamefont {Setyawan}}\ and\ \bibinfo {author} {\bibfnamefont {S.}~\bibnamefont {Curtarolo}},\ }\bibfield  {title} {\bibinfo {title} {High-throughput electronic band structure calculations: Challenges and tools},\ }\href {https://doi.org/10.1016/j.commatsci.2010.05.010} {\bibfield  {journal} {\bibinfo  {journal} {Computational Materials Science}\ }\textbf {\bibinfo {volume} {49}},\ \bibinfo {pages} {299–312} (\bibinfo {year} {2010})}\BibitemShut {NoStop}%
\bibitem [{\citenamefont {Eremin}\ \emph {et~al.}(2005)\citenamefont {Eremin}, \citenamefont {Morr}, \citenamefont {Chubukov}, \citenamefont {Bennemann},\ and\ \citenamefont {Norman}}]{Eremin2005Novel}%
  \BibitemOpen
  \bibfield  {author} {\bibinfo {author} {\bibfnamefont {I.}~\bibnamefont {Eremin}}, \bibinfo {author} {\bibfnamefont {D.~K.}\ \bibnamefont {Morr}}, \bibinfo {author} {\bibfnamefont {A.~V.}\ \bibnamefont {Chubukov}}, \bibinfo {author} {\bibfnamefont {K.~H.}\ \bibnamefont {Bennemann}},\ and\ \bibinfo {author} {\bibfnamefont {M.~R.}\ \bibnamefont {Norman}},\ }\bibfield  {title} {\bibinfo {title} {Novel neutron resonance mode in ${d}_{{x}^{2}\ensuremath{-}{y}^{2}}$-wave superconductors},\ }\href {https://doi.org/10.1103/PhysRevLett.94.147001} {\bibfield  {journal} {\bibinfo  {journal} {Phys. Rev. Lett.}\ }\textbf {\bibinfo {volume} {94}},\ \bibinfo {pages} {147001} (\bibinfo {year} {2005})}\BibitemShut {NoStop}%
\end{thebibliography}%


%apsrev4-2.bst 2019-01-14 (MD) hand-edited version of apsrev4-1.bst
%Control: key (0)
%Control: author (8) initials jnrlst
%Control: editor formatted (1) identically to author
%Control: production of article title (0) allowed
%Control: page (0) single
%Control: year (1) truncated
%Control: production of eprint (0) enabled
\begin{thebibliography}{92}%
\makeatletter
\providecommand \@ifxundefined [1]{%
 \@ifx{#1\undefined}
}%
\providecommand \@ifnum [1]{%
 \ifnum #1\expandafter \@firstoftwo
 \else \expandafter \@secondoftwo
 \fi
}%
\providecommand \@ifx [1]{%
 \ifx #1\expandafter \@firstoftwo
 \else \expandafter \@secondoftwo
 \fi
}%
\providecommand \natexlab [1]{#1}%
\providecommand \enquote  [1]{``#1''}%
\providecommand \bibnamefont  [1]{#1}%
\providecommand \bibfnamefont [1]{#1}%
\providecommand \citenamefont [1]{#1}%
\providecommand \href@noop [0]{\@secondoftwo}%
\providecommand \href [0]{\begingroup \@sanitize@url \@href}%
\providecommand \@href[1]{\@@startlink{#1}\@@href}%
\providecommand \@@href[1]{\endgroup#1\@@endlink}%
\providecommand \@sanitize@url [0]{\catcode `\\12\catcode `\$12\catcode `\&12\catcode `\#12\catcode `\^12\catcode `\_12\catcode `\%12\relax}%
\providecommand \@@startlink[1]{}%
\providecommand \@@endlink[0]{}%
\providecommand \url  [0]{\begingroup\@sanitize@url \@url }%
\providecommand \@url [1]{\endgroup\@href {#1}{\urlprefix }}%
\providecommand \urlprefix  [0]{URL }%
\providecommand \Eprint [0]{\href }%
\providecommand \doibase [0]{https://doi.org/}%
\providecommand \selectlanguage [0]{\@gobble}%
\providecommand \bibinfo  [0]{\@secondoftwo}%
\providecommand \bibfield  [0]{\@secondoftwo}%
\providecommand \translation [1]{[#1]}%
\providecommand \BibitemOpen [0]{}%
\providecommand \bibitemStop [0]{}%
\providecommand \bibitemNoStop [0]{.\EOS\space}%
\providecommand \EOS [0]{\spacefactor3000\relax}%
\providecommand \BibitemShut  [1]{\csname bibitem#1\endcsname}%
\let\auto@bib@innerbib\@empty
%</preamble>
\bibitem [{\citenamefont {Li}\ \emph {et~al.}(2019)\citenamefont {Li}, \citenamefont {Lee}, \citenamefont {Wang}, \citenamefont {Osada}, \citenamefont {Crossley}, \citenamefont {Lee}, \citenamefont {Cui}, \citenamefont {Hikita},\ and\ \citenamefont {Hwang}}]{Li2019}%
  \BibitemOpen
  \bibfield  {author} {\bibinfo {author} {\bibfnamefont {D.}~\bibnamefont {Li}}, \bibinfo {author} {\bibfnamefont {K.}~\bibnamefont {Lee}}, \bibinfo {author} {\bibfnamefont {B.~Y.}\ \bibnamefont {Wang}}, \bibinfo {author} {\bibfnamefont {M.}~\bibnamefont {Osada}}, \bibinfo {author} {\bibfnamefont {S.}~\bibnamefont {Crossley}}, \bibinfo {author} {\bibfnamefont {H.~R.}\ \bibnamefont {Lee}}, \bibinfo {author} {\bibfnamefont {Y.}~\bibnamefont {Cui}}, \bibinfo {author} {\bibfnamefont {Y.}~\bibnamefont {Hikita}},\ and\ \bibinfo {author} {\bibfnamefont {H.~Y.}\ \bibnamefont {Hwang}},\ }\bibfield  {title} {\bibinfo {title} {Superconductivity in an infinite-layer nickelate},\ }\href {https://doi.org/10.1038/s41586-019-1496-5} {\bibfield  {journal} {\bibinfo  {journal} {Nature}\ }\textbf {\bibinfo {volume} {572}},\ \bibinfo {pages} {624–627} (\bibinfo {year} {2019})}\BibitemShut {NoStop}%
\bibitem [{\citenamefont {Pan}\ \emph {et~al.}(2021)\citenamefont {Pan}, \citenamefont {Ferenc~Segedin}, \citenamefont {LaBollita}, \citenamefont {Song}, \citenamefont {Nica}, \citenamefont {Goodge}, \citenamefont {Pierce}, \citenamefont {Doyle}, \citenamefont {Novakov}, \citenamefont {Córdova~Carrizales}, \citenamefont {N’Diaye}, \citenamefont {Shafer}, \citenamefont {Paik}, \citenamefont {Heron}, \citenamefont {Mason}, \citenamefont {Yacoby}, \citenamefont {Kourkoutis}, \citenamefont {Erten}, \citenamefont {Brooks}, \citenamefont {Botana},\ and\ \citenamefont {Mundy}}]{Pan2021}%
  \BibitemOpen
  \bibfield  {author} {\bibinfo {author} {\bibfnamefont {G.~A.}\ \bibnamefont {Pan}}, \bibinfo {author} {\bibfnamefont {D.}~\bibnamefont {Ferenc~Segedin}}, \bibinfo {author} {\bibfnamefont {H.}~\bibnamefont {LaBollita}}, \bibinfo {author} {\bibfnamefont {Q.}~\bibnamefont {Song}}, \bibinfo {author} {\bibfnamefont {E.~M.}\ \bibnamefont {Nica}}, \bibinfo {author} {\bibfnamefont {B.~H.}\ \bibnamefont {Goodge}}, \bibinfo {author} {\bibfnamefont {A.~T.}\ \bibnamefont {Pierce}}, \bibinfo {author} {\bibfnamefont {S.}~\bibnamefont {Doyle}}, \bibinfo {author} {\bibfnamefont {S.}~\bibnamefont {Novakov}}, \bibinfo {author} {\bibfnamefont {D.}~\bibnamefont {Córdova~Carrizales}}, \bibinfo {author} {\bibfnamefont {A.~T.}\ \bibnamefont {N’Diaye}}, \bibinfo {author} {\bibfnamefont {P.}~\bibnamefont {Shafer}}, \bibinfo {author} {\bibfnamefont {H.}~\bibnamefont {Paik}}, \bibinfo {author} {\bibfnamefont {J.~T.}\ \bibnamefont {Heron}}, \bibinfo {author} {\bibfnamefont {J.~A.}\ \bibnamefont {Mason}}, \bibinfo {author}
  {\bibfnamefont {A.}~\bibnamefont {Yacoby}}, \bibinfo {author} {\bibfnamefont {L.~F.}\ \bibnamefont {Kourkoutis}}, \bibinfo {author} {\bibfnamefont {O.}~\bibnamefont {Erten}}, \bibinfo {author} {\bibfnamefont {C.~M.}\ \bibnamefont {Brooks}}, \bibinfo {author} {\bibfnamefont {A.~S.}\ \bibnamefont {Botana}},\ and\ \bibinfo {author} {\bibfnamefont {J.~A.}\ \bibnamefont {Mundy}},\ }\bibfield  {title} {\bibinfo {title} {Superconductivity in a quintuple-layer square-planar nickelate},\ }\href {https://doi.org/10.1038/s41563-021-01142-9} {\bibfield  {journal} {\bibinfo  {journal} {Nature Materials}\ }\textbf {\bibinfo {volume} {21}},\ \bibinfo {pages} {160–164} (\bibinfo {year} {2021})}\BibitemShut {NoStop}%
\bibitem [{\citenamefont {Osada}\ \emph {et~al.}(2020)\citenamefont {Osada}, \citenamefont {Wang}, \citenamefont {Goodge}, \citenamefont {Lee}, \citenamefont {Yoon}, \citenamefont {Sakuma}, \citenamefont {Li}, \citenamefont {Miura}, \citenamefont {Kourkoutis},\ and\ \citenamefont {Hwang}}]{Osada2020}%
  \BibitemOpen
  \bibfield  {author} {\bibinfo {author} {\bibfnamefont {M.}~\bibnamefont {Osada}}, \bibinfo {author} {\bibfnamefont {B.~Y.}\ \bibnamefont {Wang}}, \bibinfo {author} {\bibfnamefont {B.~H.}\ \bibnamefont {Goodge}}, \bibinfo {author} {\bibfnamefont {K.}~\bibnamefont {Lee}}, \bibinfo {author} {\bibfnamefont {H.}~\bibnamefont {Yoon}}, \bibinfo {author} {\bibfnamefont {K.}~\bibnamefont {Sakuma}}, \bibinfo {author} {\bibfnamefont {D.}~\bibnamefont {Li}}, \bibinfo {author} {\bibfnamefont {M.}~\bibnamefont {Miura}}, \bibinfo {author} {\bibfnamefont {L.~F.}\ \bibnamefont {Kourkoutis}},\ and\ \bibinfo {author} {\bibfnamefont {H.~Y.}\ \bibnamefont {Hwang}},\ }\bibfield  {title} {\bibinfo {title} {A superconducting praseodymium nickelate with infinite layer structure},\ }\href {https://doi.org/10.1021/acs.nanolett.0c01392} {\bibfield  {journal} {\bibinfo  {journal} {Nano Letters}\ }\textbf {\bibinfo {volume} {20}},\ \bibinfo {pages} {5735–5740} (\bibinfo {year} {2020})}\BibitemShut {NoStop}%
\bibitem [{\citenamefont {Sun}\ \emph {et~al.}(2023)\citenamefont {Sun}, \citenamefont {Huo}, \citenamefont {Hu}, \citenamefont {Li}, \citenamefont {Liu}, \citenamefont {Han}, \citenamefont {Tang}, \citenamefont {Mao}, \citenamefont {Yang}, \citenamefont {Wang} \emph {et~al.}}]{sun23}%
  \BibitemOpen
  \bibfield  {author} {\bibinfo {author} {\bibfnamefont {H.}~\bibnamefont {Sun}}, \bibinfo {author} {\bibfnamefont {M.}~\bibnamefont {Huo}}, \bibinfo {author} {\bibfnamefont {X.}~\bibnamefont {Hu}}, \bibinfo {author} {\bibfnamefont {J.}~\bibnamefont {Li}}, \bibinfo {author} {\bibfnamefont {Z.}~\bibnamefont {Liu}}, \bibinfo {author} {\bibfnamefont {Y.}~\bibnamefont {Han}}, \bibinfo {author} {\bibfnamefont {L.}~\bibnamefont {Tang}}, \bibinfo {author} {\bibfnamefont {Z.}~\bibnamefont {Mao}}, \bibinfo {author} {\bibfnamefont {P.}~\bibnamefont {Yang}}, \bibinfo {author} {\bibfnamefont {B.}~\bibnamefont {Wang}}, \emph {et~al.},\ }\bibfield  {title} {\bibinfo {title} {Signatures of superconductivity near 80 k in a nickelate under high pressure},\ }\href@noop {} {\bibfield  {journal} {\bibinfo  {journal} {Nature}\ }\textbf {\bibinfo {volume} {621}},\ \bibinfo {pages} {493} (\bibinfo {year} {2023})}\BibitemShut {NoStop}%
\bibitem [{\citenamefont {Hou}\ \emph {et~al.}(2023)\citenamefont {Hou}, \citenamefont {Yang}, \citenamefont {Liu}, \citenamefont {Li}, \citenamefont {Shan}, \citenamefont {Ma}, \citenamefont {Wang}, \citenamefont {Wang}, \citenamefont {Guo}, \citenamefont {Sun}, \citenamefont {Uwatoko}, \citenamefont {Wang}, \citenamefont {Zhang}, \citenamefont {Wang},\ and\ \citenamefont {Cheng}}]{JunHou:117302}%
  \BibitemOpen
  \bibfield  {author} {\bibinfo {author} {\bibfnamefont {J.}~\bibnamefont {Hou}}, \bibinfo {author} {\bibfnamefont {P.-T.}\ \bibnamefont {Yang}}, \bibinfo {author} {\bibfnamefont {Z.-Y.}\ \bibnamefont {Liu}}, \bibinfo {author} {\bibfnamefont {J.-Y.}\ \bibnamefont {Li}}, \bibinfo {author} {\bibfnamefont {P.-F.}\ \bibnamefont {Shan}}, \bibinfo {author} {\bibfnamefont {L.}~\bibnamefont {Ma}}, \bibinfo {author} {\bibfnamefont {G.}~\bibnamefont {Wang}}, \bibinfo {author} {\bibfnamefont {N.-N.}\ \bibnamefont {Wang}}, \bibinfo {author} {\bibfnamefont {H.-Z.}\ \bibnamefont {Guo}}, \bibinfo {author} {\bibfnamefont {J.-P.}\ \bibnamefont {Sun}}, \bibinfo {author} {\bibfnamefont {Y.}~\bibnamefont {Uwatoko}}, \bibinfo {author} {\bibfnamefont {M.}~\bibnamefont {Wang}}, \bibinfo {author} {\bibfnamefont {G.-M.}\ \bibnamefont {Zhang}}, \bibinfo {author} {\bibfnamefont {B.-S.}\ \bibnamefont {Wang}},\ and\ \bibinfo {author} {\bibfnamefont {J.-G.}\ \bibnamefont {Cheng}},\ }\bibfield  {title} {\bibinfo {title} {Emergence of
  high-temperature superconducting phase in pressurized {La$_{3}$Ni$_{2}$O$_7$} crystals},\ }\href {https://doi.org/10.1088/0256-307X/40/11/117302} {\bibfield  {journal} {\bibinfo  {journal} {Chinese Physics Letters}\ }\textbf {\bibinfo {volume} {40}},\ \bibinfo {eid} {117302} (\bibinfo {year} {2023})}\BibitemShut {NoStop}%
\bibitem [{\citenamefont {Zhang}\ \emph {et~al.}(2023{\natexlab{a}})\citenamefont {Zhang}, \citenamefont {Su}, \citenamefont {Huang}, \citenamefont {Sun}, \citenamefont {Huo}, \citenamefont {Shan}, \citenamefont {Ye}, \citenamefont {Yang}, \citenamefont {Li}, \citenamefont {Smidman} \emph {et~al.}}]{zhang2023high}%
  \BibitemOpen
  \bibfield  {author} {\bibinfo {author} {\bibfnamefont {Y.}~\bibnamefont {Zhang}}, \bibinfo {author} {\bibfnamefont {D.}~\bibnamefont {Su}}, \bibinfo {author} {\bibfnamefont {Y.}~\bibnamefont {Huang}}, \bibinfo {author} {\bibfnamefont {H.}~\bibnamefont {Sun}}, \bibinfo {author} {\bibfnamefont {M.}~\bibnamefont {Huo}}, \bibinfo {author} {\bibfnamefont {Z.}~\bibnamefont {Shan}}, \bibinfo {author} {\bibfnamefont {K.}~\bibnamefont {Ye}}, \bibinfo {author} {\bibfnamefont {Z.}~\bibnamefont {Yang}}, \bibinfo {author} {\bibfnamefont {R.}~\bibnamefont {Li}}, \bibinfo {author} {\bibfnamefont {M.}~\bibnamefont {Smidman}}, \emph {et~al.},\ }\bibfield  {title} {\bibinfo {title} {High-temperature superconductivity with zero-resistance and strange metal behavior in {La$_{3}$Ni$_{2}$O$_7$}},\ }\href@noop {} {\bibfield  {journal} {\bibinfo  {journal} {arXiv preprint arXiv:2307.14819}\ } (\bibinfo {year} {2023}{\natexlab{a}})}\BibitemShut {NoStop}%
\bibitem [{\citenamefont {Zhou}\ \emph {et~al.}(2023)\citenamefont {Zhou}, \citenamefont {Guo}, \citenamefont {Cai}, \citenamefont {Sun}, \citenamefont {Wang}, \citenamefont {Zhao}, \citenamefont {Han}, \citenamefont {Chen}, \citenamefont {Wu}, \citenamefont {Ding}, \citenamefont {Wang}, \citenamefont {Xiang}, \citenamefont {kwang Mao},\ and\ \citenamefont {Sun}}]{zhou2023evidence}%
  \BibitemOpen
  \bibfield  {author} {\bibinfo {author} {\bibfnamefont {Y.}~\bibnamefont {Zhou}}, \bibinfo {author} {\bibfnamefont {J.}~\bibnamefont {Guo}}, \bibinfo {author} {\bibfnamefont {S.}~\bibnamefont {Cai}}, \bibinfo {author} {\bibfnamefont {H.}~\bibnamefont {Sun}}, \bibinfo {author} {\bibfnamefont {P.}~\bibnamefont {Wang}}, \bibinfo {author} {\bibfnamefont {J.}~\bibnamefont {Zhao}}, \bibinfo {author} {\bibfnamefont {J.}~\bibnamefont {Han}}, \bibinfo {author} {\bibfnamefont {X.}~\bibnamefont {Chen}}, \bibinfo {author} {\bibfnamefont {Q.}~\bibnamefont {Wu}}, \bibinfo {author} {\bibfnamefont {Y.}~\bibnamefont {Ding}}, \bibinfo {author} {\bibfnamefont {M.}~\bibnamefont {Wang}}, \bibinfo {author} {\bibfnamefont {T.}~\bibnamefont {Xiang}}, \bibinfo {author} {\bibfnamefont {H.}~\bibnamefont {kwang Mao}},\ and\ \bibinfo {author} {\bibfnamefont {L.}~\bibnamefont {Sun}},\ }\href@noop {} {\bibinfo {title} {Evidence of filamentary superconductivity in pressurized {La$_{3}$Ni$_{2}$O$_7$} single crystals}} (\bibinfo {year}
  {2023}),\ \Eprint {https://arxiv.org/abs/2311.12361} {arXiv:2311.12361 [cond-mat.supr-con]} \BibitemShut {NoStop}%
\bibitem [{\citenamefont {Zhang}\ \emph {et~al.}(2023{\natexlab{b}})\citenamefont {Zhang}, \citenamefont {Pei}, \citenamefont {Wang}, \citenamefont {Zhao}, \citenamefont {Li}, \citenamefont {Cao}, \citenamefont {Zhu}, \citenamefont {Wu},\ and\ \citenamefont {Qi}}]{zhang2023effects}%
  \BibitemOpen
  \bibfield  {author} {\bibinfo {author} {\bibfnamefont {M.}~\bibnamefont {Zhang}}, \bibinfo {author} {\bibfnamefont {C.}~\bibnamefont {Pei}}, \bibinfo {author} {\bibfnamefont {Q.}~\bibnamefont {Wang}}, \bibinfo {author} {\bibfnamefont {Y.}~\bibnamefont {Zhao}}, \bibinfo {author} {\bibfnamefont {C.}~\bibnamefont {Li}}, \bibinfo {author} {\bibfnamefont {W.}~\bibnamefont {Cao}}, \bibinfo {author} {\bibfnamefont {S.}~\bibnamefont {Zhu}}, \bibinfo {author} {\bibfnamefont {J.}~\bibnamefont {Wu}},\ and\ \bibinfo {author} {\bibfnamefont {Y.}~\bibnamefont {Qi}},\ }\bibfield  {title} {\bibinfo {title} {Effects of pressure and doping on ruddlesden-popper phases},\ }\href@noop {} {\bibfield  {journal} {\bibinfo  {journal} {arXiv preprint arXiv:2309.01651}\ } (\bibinfo {year} {2023}{\natexlab{b}})}\BibitemShut {NoStop}%
\bibitem [{\citenamefont {Wang}\ \emph {et~al.}(2023{\natexlab{a}})\citenamefont {Wang}, \citenamefont {Li}, \citenamefont {Xie}, \citenamefont {Liu}, \citenamefont {Sun}, \citenamefont {Huang}, \citenamefont {Gao}, \citenamefont {Nakagawa}, \citenamefont {Fu}, \citenamefont {Dong} \emph {et~al.}}]{wang2023structure}%
  \BibitemOpen
  \bibfield  {author} {\bibinfo {author} {\bibfnamefont {L.}~\bibnamefont {Wang}}, \bibinfo {author} {\bibfnamefont {Y.}~\bibnamefont {Li}}, \bibinfo {author} {\bibfnamefont {S.}~\bibnamefont {Xie}}, \bibinfo {author} {\bibfnamefont {F.}~\bibnamefont {Liu}}, \bibinfo {author} {\bibfnamefont {H.}~\bibnamefont {Sun}}, \bibinfo {author} {\bibfnamefont {C.}~\bibnamefont {Huang}}, \bibinfo {author} {\bibfnamefont {Y.}~\bibnamefont {Gao}}, \bibinfo {author} {\bibfnamefont {T.}~\bibnamefont {Nakagawa}}, \bibinfo {author} {\bibfnamefont {B.}~\bibnamefont {Fu}}, \bibinfo {author} {\bibfnamefont {B.}~\bibnamefont {Dong}}, \emph {et~al.},\ }\bibfield  {title} {\bibinfo {title} {Structure responsible for the superconducting state in {La$_{3}$Ni$_{2}$O$_7$} at low temperature and high pressure conditions},\ }\href@noop {} {\bibfield  {journal} {\bibinfo  {journal} {arXiv preprint arXiv:2311.09186}\ } (\bibinfo {year} {2023}{\natexlab{a}})}\BibitemShut {NoStop}%
\bibitem [{\citenamefont {Wang}\ \emph {et~al.}(2023{\natexlab{b}})\citenamefont {Wang}, \citenamefont {Wang}, \citenamefont {Hou}, \citenamefont {Ma}, \citenamefont {Shi}, \citenamefont {Ren}, \citenamefont {Gu}, \citenamefont {Shen}, \citenamefont {Ma}, \citenamefont {Yang} \emph {et~al.}}]{wang2023pressure}%
  \BibitemOpen
  \bibfield  {author} {\bibinfo {author} {\bibfnamefont {G.}~\bibnamefont {Wang}}, \bibinfo {author} {\bibfnamefont {N.}~\bibnamefont {Wang}}, \bibinfo {author} {\bibfnamefont {J.}~\bibnamefont {Hou}}, \bibinfo {author} {\bibfnamefont {L.}~\bibnamefont {Ma}}, \bibinfo {author} {\bibfnamefont {L.}~\bibnamefont {Shi}}, \bibinfo {author} {\bibfnamefont {Z.}~\bibnamefont {Ren}}, \bibinfo {author} {\bibfnamefont {Y.}~\bibnamefont {Gu}}, \bibinfo {author} {\bibfnamefont {X.}~\bibnamefont {Shen}}, \bibinfo {author} {\bibfnamefont {H.}~\bibnamefont {Ma}}, \bibinfo {author} {\bibfnamefont {P.}~\bibnamefont {Yang}}, \emph {et~al.},\ }\bibfield  {title} {\bibinfo {title} {Pressure-induced superconductivity in polycrystalline {La$_{3}$Ni$_{2}$O$_7$}},\ }\href@noop {} {\bibfield  {journal} {\bibinfo  {journal} {arXiv preprint arXiv:2309.17378}\ } (\bibinfo {year} {2023}{\natexlab{b}})}\BibitemShut {NoStop}%
\bibitem [{\citenamefont {Dong}\ \emph {et~al.}(2023)\citenamefont {Dong}, \citenamefont {Huo}, \citenamefont {Li}, \citenamefont {Li}, \citenamefont {Li}, \citenamefont {Sun}, \citenamefont {Lu}, \citenamefont {Wang}, \citenamefont {Wang},\ and\ \citenamefont {Chen}}]{dong2023visualization}%
  \BibitemOpen
  \bibfield  {author} {\bibinfo {author} {\bibfnamefont {Z.}~\bibnamefont {Dong}}, \bibinfo {author} {\bibfnamefont {M.}~\bibnamefont {Huo}}, \bibinfo {author} {\bibfnamefont {J.}~\bibnamefont {Li}}, \bibinfo {author} {\bibfnamefont {J.}~\bibnamefont {Li}}, \bibinfo {author} {\bibfnamefont {P.}~\bibnamefont {Li}}, \bibinfo {author} {\bibfnamefont {H.}~\bibnamefont {Sun}}, \bibinfo {author} {\bibfnamefont {Y.}~\bibnamefont {Lu}}, \bibinfo {author} {\bibfnamefont {M.}~\bibnamefont {Wang}}, \bibinfo {author} {\bibfnamefont {Y.}~\bibnamefont {Wang}},\ and\ \bibinfo {author} {\bibfnamefont {Z.}~\bibnamefont {Chen}},\ }\bibfield  {title} {\bibinfo {title} {Visualization of oxygen vacancies and self-doped ligand holes in {La$_{3}$Ni$_{2}$O$_{7-\delta}$}},\ }\href@noop {} {\bibfield  {journal} {\bibinfo  {journal} {arXiv preprint arXiv:2312.15727}\ } (\bibinfo {year} {2023})}\BibitemShut {NoStop}%
\bibitem [{\citenamefont {Sakakibara}\ \emph {et~al.}(2023)\citenamefont {Sakakibara}, \citenamefont {Ochi}, \citenamefont {Nagata}, \citenamefont {Ueki}, \citenamefont {Sakurai}, \citenamefont {Matsumoto}, \citenamefont {Terashima}, \citenamefont {Hirose}, \citenamefont {Ohta}, \citenamefont {Kato}, \citenamefont {Takano},\ and\ \citenamefont {Kuroki}}]{sakakibara2023theoretical}%
  \BibitemOpen
  \bibfield  {author} {\bibinfo {author} {\bibfnamefont {H.}~\bibnamefont {Sakakibara}}, \bibinfo {author} {\bibfnamefont {M.}~\bibnamefont {Ochi}}, \bibinfo {author} {\bibfnamefont {H.}~\bibnamefont {Nagata}}, \bibinfo {author} {\bibfnamefont {Y.}~\bibnamefont {Ueki}}, \bibinfo {author} {\bibfnamefont {H.}~\bibnamefont {Sakurai}}, \bibinfo {author} {\bibfnamefont {R.}~\bibnamefont {Matsumoto}}, \bibinfo {author} {\bibfnamefont {K.}~\bibnamefont {Terashima}}, \bibinfo {author} {\bibfnamefont {K.}~\bibnamefont {Hirose}}, \bibinfo {author} {\bibfnamefont {H.}~\bibnamefont {Ohta}}, \bibinfo {author} {\bibfnamefont {M.}~\bibnamefont {Kato}}, \bibinfo {author} {\bibfnamefont {Y.}~\bibnamefont {Takano}},\ and\ \bibinfo {author} {\bibfnamefont {K.}~\bibnamefont {Kuroki}},\ }\href@noop {} {\bibinfo {title} {Theoretical analysis on the possibility of superconductivity in a trilayer ruddlesden-popper nickelate {La$_4$Ni$_3$O$_{10}$} under pressure and its experimental examination: comparison with
  {La$_{3}$Ni$_{2}$O$_7$}}} (\bibinfo {year} {2023}),\ \Eprint {https://arxiv.org/abs/2309.09462} {arXiv:2309.09462 [cond-mat.supr-con]} \BibitemShut {NoStop}%
\bibitem [{\citenamefont {Li}\ \emph {et~al.}(2023)\citenamefont {Li}, \citenamefont {Zhang}, \citenamefont {Xiang}, \citenamefont {Zhang}, \citenamefont {Zhu},\ and\ \citenamefont {Wen}}]{li2023signature}%
  \BibitemOpen
  \bibfield  {author} {\bibinfo {author} {\bibfnamefont {Q.}~\bibnamefont {Li}}, \bibinfo {author} {\bibfnamefont {Y.-J.}\ \bibnamefont {Zhang}}, \bibinfo {author} {\bibfnamefont {Z.-N.}\ \bibnamefont {Xiang}}, \bibinfo {author} {\bibfnamefont {Y.}~\bibnamefont {Zhang}}, \bibinfo {author} {\bibfnamefont {X.}~\bibnamefont {Zhu}},\ and\ \bibinfo {author} {\bibfnamefont {H.-H.}\ \bibnamefont {Wen}},\ }\href@noop {} {\bibinfo {title} {Signature of superconductivity in pressurized {La$_4$Ni$_3$O$_{10}$}}} (\bibinfo {year} {2023}),\ \Eprint {https://arxiv.org/abs/2311.05453} {arXiv:2311.05453 [cond-mat.supr-con]} \BibitemShut {NoStop}%
\bibitem [{\citenamefont {Zhang}\ \emph {et~al.}(2023{\natexlab{c}})\citenamefont {Zhang}, \citenamefont {Pei}, \citenamefont {Du}, \citenamefont {Cao}, \citenamefont {Wang}, \citenamefont {Wu}, \citenamefont {Li}, \citenamefont {Zhao}, \citenamefont {Li}, \citenamefont {Cao}, \citenamefont {Zhu}, \citenamefont {Zhang}, \citenamefont {Yu}, \citenamefont {Cheng}, \citenamefont {Zhao}, \citenamefont {Chen}, \citenamefont {Guo}, \citenamefont {Yang},\ and\ \citenamefont {Qi}}]{zhang2023superconductivity}%
  \BibitemOpen
  \bibfield  {author} {\bibinfo {author} {\bibfnamefont {M.}~\bibnamefont {Zhang}}, \bibinfo {author} {\bibfnamefont {C.}~\bibnamefont {Pei}}, \bibinfo {author} {\bibfnamefont {X.}~\bibnamefont {Du}}, \bibinfo {author} {\bibfnamefont {Y.}~\bibnamefont {Cao}}, \bibinfo {author} {\bibfnamefont {Q.}~\bibnamefont {Wang}}, \bibinfo {author} {\bibfnamefont {J.}~\bibnamefont {Wu}}, \bibinfo {author} {\bibfnamefont {Y.}~\bibnamefont {Li}}, \bibinfo {author} {\bibfnamefont {Y.}~\bibnamefont {Zhao}}, \bibinfo {author} {\bibfnamefont {C.}~\bibnamefont {Li}}, \bibinfo {author} {\bibfnamefont {W.}~\bibnamefont {Cao}}, \bibinfo {author} {\bibfnamefont {S.}~\bibnamefont {Zhu}}, \bibinfo {author} {\bibfnamefont {Q.}~\bibnamefont {Zhang}}, \bibinfo {author} {\bibfnamefont {N.}~\bibnamefont {Yu}}, \bibinfo {author} {\bibfnamefont {P.}~\bibnamefont {Cheng}}, \bibinfo {author} {\bibfnamefont {J.}~\bibnamefont {Zhao}}, \bibinfo {author} {\bibfnamefont {Y.}~\bibnamefont {Chen}}, \bibinfo {author} {\bibfnamefont {H.}~\bibnamefont
  {Guo}}, \bibinfo {author} {\bibfnamefont {L.}~\bibnamefont {Yang}},\ and\ \bibinfo {author} {\bibfnamefont {Y.}~\bibnamefont {Qi}},\ }\href@noop {} {\bibinfo {title} {Superconductivity in trilayer nickelate {La$_4$Ni$_3$O$_{10}$} under pressure}} (\bibinfo {year} {2023}{\natexlab{c}}),\ \Eprint {https://arxiv.org/abs/2311.07423} {arXiv:2311.07423 [cond-mat.supr-con]} \BibitemShut {NoStop}%
\bibitem [{\citenamefont {Zhu}\ \emph {et~al.}(2024)\citenamefont {Zhu}, \citenamefont {Zhang}, \citenamefont {Pan}, \citenamefont {Chen}, \citenamefont {Peng}, \citenamefont {Chen}, \citenamefont {Ren}, \citenamefont {Liu}, \citenamefont {Li}, \citenamefont {Xing}, \citenamefont {Han}, \citenamefont {Wang}, \citenamefont {Jia}, \citenamefont {Wo}, \citenamefont {Gu}, \citenamefont {Gu}, \citenamefont {Ji}, \citenamefont {Wang}, \citenamefont {Gou}, \citenamefont {Shen}, \citenamefont {Ying}, \citenamefont {Chen}, \citenamefont {Yang}, \citenamefont {Zheng}, \citenamefont {Zeng}, \citenamefont {gang Guo},\ and\ \citenamefont {Zhao}}]{zhu2024superconductivity}%
  \BibitemOpen
  \bibfield  {author} {\bibinfo {author} {\bibfnamefont {Y.}~\bibnamefont {Zhu}}, \bibinfo {author} {\bibfnamefont {E.}~\bibnamefont {Zhang}}, \bibinfo {author} {\bibfnamefont {B.}~\bibnamefont {Pan}}, \bibinfo {author} {\bibfnamefont {X.}~\bibnamefont {Chen}}, \bibinfo {author} {\bibfnamefont {D.}~\bibnamefont {Peng}}, \bibinfo {author} {\bibfnamefont {L.}~\bibnamefont {Chen}}, \bibinfo {author} {\bibfnamefont {H.}~\bibnamefont {Ren}}, \bibinfo {author} {\bibfnamefont {F.}~\bibnamefont {Liu}}, \bibinfo {author} {\bibfnamefont {N.}~\bibnamefont {Li}}, \bibinfo {author} {\bibfnamefont {Z.}~\bibnamefont {Xing}}, \bibinfo {author} {\bibfnamefont {J.}~\bibnamefont {Han}}, \bibinfo {author} {\bibfnamefont {J.}~\bibnamefont {Wang}}, \bibinfo {author} {\bibfnamefont {D.}~\bibnamefont {Jia}}, \bibinfo {author} {\bibfnamefont {H.}~\bibnamefont {Wo}}, \bibinfo {author} {\bibfnamefont {Y.}~\bibnamefont {Gu}}, \bibinfo {author} {\bibfnamefont {Y.}~\bibnamefont {Gu}}, \bibinfo {author} {\bibfnamefont {L.}~\bibnamefont
  {Ji}}, \bibinfo {author} {\bibfnamefont {W.}~\bibnamefont {Wang}}, \bibinfo {author} {\bibfnamefont {H.}~\bibnamefont {Gou}}, \bibinfo {author} {\bibfnamefont {Y.}~\bibnamefont {Shen}}, \bibinfo {author} {\bibfnamefont {T.}~\bibnamefont {Ying}}, \bibinfo {author} {\bibfnamefont {X.}~\bibnamefont {Chen}}, \bibinfo {author} {\bibfnamefont {W.}~\bibnamefont {Yang}}, \bibinfo {author} {\bibfnamefont {C.}~\bibnamefont {Zheng}}, \bibinfo {author} {\bibfnamefont {Q.}~\bibnamefont {Zeng}}, \bibinfo {author} {\bibfnamefont {J.}~\bibnamefont {gang Guo}},\ and\ \bibinfo {author} {\bibfnamefont {J.}~\bibnamefont {Zhao}},\ }\href@noop {} {\bibinfo {title} {Superconductivity in trilayer nickelate {La$_4$Ni$_3$O$_{10}$} single crystals}} (\bibinfo {year} {2024}),\ \Eprint {https://arxiv.org/abs/2311.07353} {arXiv:2311.07353 [cond-mat.supr-con]} \BibitemShut {NoStop}%
\bibitem [{\citenamefont {Lechermann}\ \emph {et~al.}(2023)\citenamefont {Lechermann}, \citenamefont {Gondolf}, \citenamefont {B\"otzel},\ and\ \citenamefont {Eremin}}]{lechermann2023electronic}%
  \BibitemOpen
  \bibfield  {author} {\bibinfo {author} {\bibfnamefont {F.}~\bibnamefont {Lechermann}}, \bibinfo {author} {\bibfnamefont {J.}~\bibnamefont {Gondolf}}, \bibinfo {author} {\bibfnamefont {S.}~\bibnamefont {B\"otzel}},\ and\ \bibinfo {author} {\bibfnamefont {I.~M.}\ \bibnamefont {Eremin}},\ }\bibfield  {title} {\bibinfo {title} {Electronic correlations and superconducting instability in {${\mathrm{La}}_{3}{\mathrm{Ni}}_{2}{\mathrm{O}}_{7}$} under high pressure},\ }\href {https://doi.org/10.1103/PhysRevB.108.L201121} {\bibfield  {journal} {\bibinfo  {journal} {Phys. Rev. B}\ }\textbf {\bibinfo {volume} {108}},\ \bibinfo {pages} {L201121} (\bibinfo {year} {2023})}\BibitemShut {NoStop}%
\bibitem [{\citenamefont {Liu}\ \emph {et~al.}(2023{\natexlab{a}})\citenamefont {Liu}, \citenamefont {Xia}, \citenamefont {Zhou},\ and\ \citenamefont {Chen}}]{liu2023role}%
  \BibitemOpen
  \bibfield  {author} {\bibinfo {author} {\bibfnamefont {H.}~\bibnamefont {Liu}}, \bibinfo {author} {\bibfnamefont {C.}~\bibnamefont {Xia}}, \bibinfo {author} {\bibfnamefont {S.}~\bibnamefont {Zhou}},\ and\ \bibinfo {author} {\bibfnamefont {H.}~\bibnamefont {Chen}},\ }\bibfield  {title} {\bibinfo {title} {Role of crystal-field-splitting and longe-range-hoppings on superconducting pairing symmetry of {La$_{3}$Ni$_{2}$O$_7$}},\ }\href@noop {} {\bibfield  {journal} {\bibinfo  {journal} {arXiv preprint arXiv:2311.07316}\ } (\bibinfo {year} {2023}{\natexlab{a}})}\BibitemShut {NoStop}%
\bibitem [{\citenamefont {Oh}\ and\ \citenamefont {Zhang}(2023)}]{oh2023type}%
  \BibitemOpen
  \bibfield  {author} {\bibinfo {author} {\bibfnamefont {H.}~\bibnamefont {Oh}}\ and\ \bibinfo {author} {\bibfnamefont {Y.-H.}\ \bibnamefont {Zhang}},\ }\bibfield  {title} {\bibinfo {title} {Type ii tj model and shared antiferromagnetic spin coupling from hund's rule in superconducting {La$_{3}$Ni$_{2}$O$_7$}},\ }\href@noop {} {\bibfield  {journal} {\bibinfo  {journal} {arXiv preprint arXiv:2307.15706}\ } (\bibinfo {year} {2023})}\BibitemShut {NoStop}%
\bibitem [{\citenamefont {Luo}\ \emph {et~al.}(2023{\natexlab{a}})\citenamefont {Luo}, \citenamefont {Lv}, \citenamefont {Wang}, \citenamefont {W{\'u}},\ and\ \citenamefont {Yao}}]{luo2023high}%
  \BibitemOpen
  \bibfield  {author} {\bibinfo {author} {\bibfnamefont {Z.}~\bibnamefont {Luo}}, \bibinfo {author} {\bibfnamefont {B.}~\bibnamefont {Lv}}, \bibinfo {author} {\bibfnamefont {M.}~\bibnamefont {Wang}}, \bibinfo {author} {\bibfnamefont {W.}~\bibnamefont {W{\'u}}},\ and\ \bibinfo {author} {\bibfnamefont {D.-x.}\ \bibnamefont {Yao}},\ }\bibfield  {title} {\bibinfo {title} {High-{$T_c$} superconductivity in {La$_{3}$Ni$_{2}$O$_7$} based on the bilayer two-orbital tj model},\ }\href@noop {} {\bibfield  {journal} {\bibinfo  {journal} {arXiv preprint arXiv:2308.16564}\ } (\bibinfo {year} {2023}{\natexlab{a}})}\BibitemShut {NoStop}%
\bibitem [{\citenamefont {Qin}\ and\ \citenamefont {Yang}(2023)}]{qin2023high}%
  \BibitemOpen
  \bibfield  {author} {\bibinfo {author} {\bibfnamefont {Q.}~\bibnamefont {Qin}}\ and\ \bibinfo {author} {\bibfnamefont {Y.-f.}\ \bibnamefont {Yang}},\ }\bibfield  {title} {\bibinfo {title} {High-{T$_c$} superconductivity by mobilizing local spin singlets and possible route to higher {T$_c$} in pressurized {La$_{3}$Ni$_{2}$O$_7$}},\ }\href@noop {} {\bibfield  {journal} {\bibinfo  {journal} {Physical Review B}\ }\textbf {\bibinfo {volume} {108}},\ \bibinfo {pages} {L140504} (\bibinfo {year} {2023})}\BibitemShut {NoStop}%
\bibitem [{\citenamefont {Huang}\ \emph {et~al.}(2023)\citenamefont {Huang}, \citenamefont {Wang},\ and\ \citenamefont {Zhou}}]{huang2023impurity}%
  \BibitemOpen
  \bibfield  {author} {\bibinfo {author} {\bibfnamefont {J.}~\bibnamefont {Huang}}, \bibinfo {author} {\bibfnamefont {Z.}~\bibnamefont {Wang}},\ and\ \bibinfo {author} {\bibfnamefont {T.}~\bibnamefont {Zhou}},\ }\bibfield  {title} {\bibinfo {title} {Impurity and vortex states in the bilayer high-temperature superconductor {La$_{3}$Ni$_{2}$O$_7$}},\ }\href@noop {} {\bibfield  {journal} {\bibinfo  {journal} {Physical Review B}\ }\textbf {\bibinfo {volume} {108}},\ \bibinfo {pages} {174501} (\bibinfo {year} {2023})}\BibitemShut {NoStop}%
\bibitem [{\citenamefont {Qu}\ \emph {et~al.}(2024)\citenamefont {Qu}, \citenamefont {Qu}, \citenamefont {Chen}, \citenamefont {Wu}, \citenamefont {Yang}, \citenamefont {Li},\ and\ \citenamefont {Su}}]{qu2023bilayer}%
  \BibitemOpen
  \bibfield  {author} {\bibinfo {author} {\bibfnamefont {X.-Z.}\ \bibnamefont {Qu}}, \bibinfo {author} {\bibfnamefont {D.-W.}\ \bibnamefont {Qu}}, \bibinfo {author} {\bibfnamefont {J.}~\bibnamefont {Chen}}, \bibinfo {author} {\bibfnamefont {C.}~\bibnamefont {Wu}}, \bibinfo {author} {\bibfnamefont {F.}~\bibnamefont {Yang}}, \bibinfo {author} {\bibfnamefont {W.}~\bibnamefont {Li}},\ and\ \bibinfo {author} {\bibfnamefont {G.}~\bibnamefont {Su}},\ }\bibfield  {title} {\bibinfo {title} {Bilayer {$t$-$J$-$J_{\perp}$} model and magnetically mediated pairing in the pressurized nickelate {{La$_{3}$Ni$_{2}$O$_{7}$}}},\ }\href@noop {} {\bibfield  {journal} {\bibinfo  {journal} {Physical Review Letters}\ }\textbf {\bibinfo {volume} {132}},\ \bibinfo {pages} {036502} (\bibinfo {year} {2024})}\BibitemShut {NoStop}%
\bibitem [{\citenamefont {Zhang}\ \emph {et~al.}(2023{\natexlab{d}})\citenamefont {Zhang}, \citenamefont {Lin}, \citenamefont {Moreo}, \citenamefont {Maier},\ and\ \citenamefont {Dagotto}}]{zhang2023trends}%
  \BibitemOpen
  \bibfield  {author} {\bibinfo {author} {\bibfnamefont {Y.}~\bibnamefont {Zhang}}, \bibinfo {author} {\bibfnamefont {L.-F.}\ \bibnamefont {Lin}}, \bibinfo {author} {\bibfnamefont {A.}~\bibnamefont {Moreo}}, \bibinfo {author} {\bibfnamefont {T.~A.}\ \bibnamefont {Maier}},\ and\ \bibinfo {author} {\bibfnamefont {E.}~\bibnamefont {Dagotto}},\ }\bibfield  {title} {\bibinfo {title} {Trends in electronic structures and s$\pm$-wave pairing for the rare-earth series in bilayer nickelate superconductor {R$_3$Ni$_2$O$_7$}},\ }\href@noop {} {\bibfield  {journal} {\bibinfo  {journal} {Physical Review B}\ }\textbf {\bibinfo {volume} {108}},\ \bibinfo {pages} {165141} (\bibinfo {year} {2023}{\natexlab{d}})}\BibitemShut {NoStop}%
\bibitem [{\citenamefont {Yang}\ \emph {et~al.}(2023{\natexlab{a}})\citenamefont {Yang}, \citenamefont {Wang},\ and\ \citenamefont {Wang}}]{yang2023possible}%
  \BibitemOpen
  \bibfield  {author} {\bibinfo {author} {\bibfnamefont {Q.-G.}\ \bibnamefont {Yang}}, \bibinfo {author} {\bibfnamefont {D.}~\bibnamefont {Wang}},\ and\ \bibinfo {author} {\bibfnamefont {Q.-H.}\ \bibnamefont {Wang}},\ }\bibfield  {title} {\bibinfo {title} {Possible ${s}_{\ifmmode\pm\else\textpm\fi{}}$-wave superconductivity in {${\mathrm{La}}_{3}{\mathrm{Ni}}_{2}{\mathrm{O}}_{7}$}},\ }\href {https://doi.org/10.1103/PhysRevB.108.L140505} {\bibfield  {journal} {\bibinfo  {journal} {Phys. Rev. B}\ }\textbf {\bibinfo {volume} {108}},\ \bibinfo {pages} {L140505} (\bibinfo {year} {2023}{\natexlab{a}})}\BibitemShut {NoStop}%
\bibitem [{\citenamefont {Zhang}\ \emph {et~al.}(2023{\natexlab{e}})\citenamefont {Zhang}, \citenamefont {Lin}, \citenamefont {Moreo}, \citenamefont {Maier},\ and\ \citenamefont {Dagotto}}]{zhang2023structural}%
  \BibitemOpen
  \bibfield  {author} {\bibinfo {author} {\bibfnamefont {Y.}~\bibnamefont {Zhang}}, \bibinfo {author} {\bibfnamefont {L.-F.}\ \bibnamefont {Lin}}, \bibinfo {author} {\bibfnamefont {A.}~\bibnamefont {Moreo}}, \bibinfo {author} {\bibfnamefont {T.~A.}\ \bibnamefont {Maier}},\ and\ \bibinfo {author} {\bibfnamefont {E.}~\bibnamefont {Dagotto}},\ }\bibfield  {title} {\bibinfo {title} {Structural phase transition, $s\pm$-wave pairing and magnetic stripe order in the bilayered nickelate superconductor {La$_{3}$Ni$_{2}$O$_7$} under pressure},\ }\href@noop {} {\bibfield  {journal} {\bibinfo  {journal} {arXiv preprint arXiv:2307.15276}\ } (\bibinfo {year} {2023}{\natexlab{e}})}\BibitemShut {NoStop}%
\bibitem [{\citenamefont {Yang}\ \emph {et~al.}(2023{\natexlab{b}})\citenamefont {Yang}, \citenamefont {Oh},\ and\ \citenamefont {Zhang}}]{yang2023strong}%
  \BibitemOpen
  \bibfield  {author} {\bibinfo {author} {\bibfnamefont {H.}~\bibnamefont {Yang}}, \bibinfo {author} {\bibfnamefont {H.}~\bibnamefont {Oh}},\ and\ \bibinfo {author} {\bibfnamefont {Y.-H.}\ \bibnamefont {Zhang}},\ }\bibfield  {title} {\bibinfo {title} {Strong pairing from doping-induced feshbach resonance and second fermi liquid through doping a bilayer spin-one mott insulator: application to {{La$_{3}$Ni$_{2}$O$_{7}$}}},\ }\href@noop {} {\bibfield  {journal} {\bibinfo  {journal} {arXiv preprint arXiv:2309.15095}\ } (\bibinfo {year} {2023}{\natexlab{b}})}\BibitemShut {NoStop}%
\bibitem [{\citenamefont {Heier}\ \emph {et~al.}(2023)\citenamefont {Heier}, \citenamefont {Park},\ and\ \citenamefont {Savrasov}}]{heier2023competing}%
  \BibitemOpen
  \bibfield  {author} {\bibinfo {author} {\bibfnamefont {G.}~\bibnamefont {Heier}}, \bibinfo {author} {\bibfnamefont {K.}~\bibnamefont {Park}},\ and\ \bibinfo {author} {\bibfnamefont {S.~Y.}\ \bibnamefont {Savrasov}},\ }\href@noop {} {\bibinfo {title} {Competing d$_{xy}$ and s$_{\pm }$ pairing symmetries in superconducting {La$_{3}$Ni$_{2}$O$_{7}$} emerge from lda+flex calculations}} (\bibinfo {year} {2023}),\ \Eprint {https://arxiv.org/abs/2312.04401} {arXiv:2312.04401 [cond-mat.supr-con]} \BibitemShut {NoStop}%
\bibitem [{\citenamefont {Jiang}\ \emph {et~al.}(2023)\citenamefont {Jiang}, \citenamefont {Wang},\ and\ \citenamefont {Zhang}}]{jiang2023high}%
  \BibitemOpen
  \bibfield  {author} {\bibinfo {author} {\bibfnamefont {K.}~\bibnamefont {Jiang}}, \bibinfo {author} {\bibfnamefont {Z.}~\bibnamefont {Wang}},\ and\ \bibinfo {author} {\bibfnamefont {F.-C.}\ \bibnamefont {Zhang}},\ }\bibfield  {title} {\bibinfo {title} {High temperature superconductivity in {La$_{3}$Ni$_{2}$O$_{7}$}},\ }\href@noop {} {\bibfield  {journal} {\bibinfo  {journal} {Chinese Physics Letters}\ } (\bibinfo {year} {2023})}\BibitemShut {NoStop}%
\bibitem [{\citenamefont {Ryee}\ \emph {et~al.}(2023)\citenamefont {Ryee}, \citenamefont {Witt},\ and\ \citenamefont {Wehling}}]{ryee2023critical}%
  \BibitemOpen
  \bibfield  {author} {\bibinfo {author} {\bibfnamefont {S.}~\bibnamefont {Ryee}}, \bibinfo {author} {\bibfnamefont {N.}~\bibnamefont {Witt}},\ and\ \bibinfo {author} {\bibfnamefont {T.~O.}\ \bibnamefont {Wehling}},\ }\bibfield  {title} {\bibinfo {title} {Critical role of interlayer dimer correlations in the superconductivity of {La$_{3}$Ni$_{2}$O$_7$}},\ }\href@noop {} {\bibfield  {journal} {\bibinfo  {journal} {arXiv preprint arXiv:2310.17465}\ } (\bibinfo {year} {2023})}\BibitemShut {NoStop}%
\bibitem [{\citenamefont {Tian}\ \emph {et~al.}(2023)\citenamefont {Tian}, \citenamefont {Chen}, \citenamefont {Wang}, \citenamefont {He},\ and\ \citenamefont {Lu}}]{tian2023correlation}%
  \BibitemOpen
  \bibfield  {author} {\bibinfo {author} {\bibfnamefont {Y.-H.}\ \bibnamefont {Tian}}, \bibinfo {author} {\bibfnamefont {Y.}~\bibnamefont {Chen}}, \bibinfo {author} {\bibfnamefont {J.-M.}\ \bibnamefont {Wang}}, \bibinfo {author} {\bibfnamefont {R.-Q.}\ \bibnamefont {He}},\ and\ \bibinfo {author} {\bibfnamefont {Z.-Y.}\ \bibnamefont {Lu}},\ }\bibfield  {title} {\bibinfo {title} {Correlation effects and concomitant two-orbital $s_{\pm}$-wave superconductivity in la {La$_{3}$Ni$_{2}$O$_7$} under high pressure},\ }\href@noop {} {\bibfield  {journal} {\bibinfo  {journal} {arXiv preprint arXiv:2308.09698}\ } (\bibinfo {year} {2023})}\BibitemShut {NoStop}%
\bibitem [{\citenamefont {Liao}\ \emph {et~al.}(2023)\citenamefont {Liao}, \citenamefont {Chen}, \citenamefont {Duan}, \citenamefont {Wang}, \citenamefont {Liu}, \citenamefont {Yu},\ and\ \citenamefont {Si}}]{liao2023electron}%
  \BibitemOpen
  \bibfield  {author} {\bibinfo {author} {\bibfnamefont {Z.}~\bibnamefont {Liao}}, \bibinfo {author} {\bibfnamefont {L.}~\bibnamefont {Chen}}, \bibinfo {author} {\bibfnamefont {G.}~\bibnamefont {Duan}}, \bibinfo {author} {\bibfnamefont {Y.}~\bibnamefont {Wang}}, \bibinfo {author} {\bibfnamefont {C.}~\bibnamefont {Liu}}, \bibinfo {author} {\bibfnamefont {R.}~\bibnamefont {Yu}},\ and\ \bibinfo {author} {\bibfnamefont {Q.}~\bibnamefont {Si}},\ }\bibfield  {title} {\bibinfo {title} {Electron correlations and superconductivity in {{La$_{3}$Ni$_{2}$O$_{7}$}} under pressure tuning},\ }\href@noop {} {\bibfield  {journal} {\bibinfo  {journal} {arXiv preprint arXiv:2307.16697}\ } (\bibinfo {year} {2023})}\BibitemShut {NoStop}%
\bibitem [{\citenamefont {Kaneko}\ \emph {et~al.}(2023)\citenamefont {Kaneko}, \citenamefont {Sakakibara}, \citenamefont {Ochi},\ and\ \citenamefont {Kuroki}}]{kaneko2023pair}%
  \BibitemOpen
  \bibfield  {author} {\bibinfo {author} {\bibfnamefont {T.}~\bibnamefont {Kaneko}}, \bibinfo {author} {\bibfnamefont {H.}~\bibnamefont {Sakakibara}}, \bibinfo {author} {\bibfnamefont {M.}~\bibnamefont {Ochi}},\ and\ \bibinfo {author} {\bibfnamefont {K.}~\bibnamefont {Kuroki}},\ }\bibfield  {title} {\bibinfo {title} {Pair correlations in the two-orbital hubbard ladder: Implications on superconductivity in the bilayer nickelate},\ }\href@noop {} {\bibfield  {journal} {\bibinfo  {journal} {arXiv preprint arXiv:2310.01952}\ } (\bibinfo {year} {2023})}\BibitemShut {NoStop}%
\bibitem [{\citenamefont {Luo}\ \emph {et~al.}(2023{\natexlab{b}})\citenamefont {Luo}, \citenamefont {Hu}, \citenamefont {Wang}, \citenamefont {W\'u},\ and\ \citenamefont {Yao}}]{LuoModel}%
  \BibitemOpen
  \bibfield  {author} {\bibinfo {author} {\bibfnamefont {Z.}~\bibnamefont {Luo}}, \bibinfo {author} {\bibfnamefont {X.}~\bibnamefont {Hu}}, \bibinfo {author} {\bibfnamefont {M.}~\bibnamefont {Wang}}, \bibinfo {author} {\bibfnamefont {W.}~\bibnamefont {W\'u}},\ and\ \bibinfo {author} {\bibfnamefont {D.-X.}\ \bibnamefont {Yao}},\ }\bibfield  {title} {\bibinfo {title} {Bilayer two-orbital model of {$\mathrm{L}{\mathrm{a}}_{3}\mathrm{N}{\mathrm{i}}_{2}{\mathrm{O}}_{7}$} under pressure},\ }\href {https://doi.org/10.1103/PhysRevLett.131.126001} {\bibfield  {journal} {\bibinfo  {journal} {Phys. Rev. Lett.}\ }\textbf {\bibinfo {volume} {131}},\ \bibinfo {pages} {126001} (\bibinfo {year} {2023}{\natexlab{b}})}\BibitemShut {NoStop}%
\bibitem [{\citenamefont {Chen}\ \emph {et~al.}(2023{\natexlab{a}})\citenamefont {Chen}, \citenamefont {Yang},\ and\ \citenamefont {Li}}]{chen2023orbital}%
  \BibitemOpen
  \bibfield  {author} {\bibinfo {author} {\bibfnamefont {J.}~\bibnamefont {Chen}}, \bibinfo {author} {\bibfnamefont {F.}~\bibnamefont {Yang}},\ and\ \bibinfo {author} {\bibfnamefont {W.}~\bibnamefont {Li}},\ }\bibfield  {title} {\bibinfo {title} {Orbital-selective superconductivity in the pressurized bilayer nickelate {{La$_{3}$Ni$_{2}$O$_{7}$}} : An infinite projected entangled-pair state study},\ }\href@noop {} {\bibfield  {journal} {\bibinfo  {journal} {arXiv preprint arXiv:2311.05491}\ } (\bibinfo {year} {2023}{\natexlab{a}})}\BibitemShut {NoStop}%
\bibitem [{\citenamefont {Shen}\ \emph {et~al.}(2023)\citenamefont {Shen}, \citenamefont {Qin},\ and\ \citenamefont {Zhang}}]{shen2023effective}%
  \BibitemOpen
  \bibfield  {author} {\bibinfo {author} {\bibfnamefont {Y.}~\bibnamefont {Shen}}, \bibinfo {author} {\bibfnamefont {M.}~\bibnamefont {Qin}},\ and\ \bibinfo {author} {\bibfnamefont {G.-M.}\ \bibnamefont {Zhang}},\ }\bibfield  {title} {\bibinfo {title} {Effective bi-layer model hamiltonian and density-matrix renormalization group study for the high-tc superconductivity in {La$_{3}$Ni$_{2}$O$_7$} under high pressure},\ }\href@noop {} {\bibfield  {journal} {\bibinfo  {journal} {arXiv preprint arXiv:2306.07837}\ } (\bibinfo {year} {2023})}\BibitemShut {NoStop}%
\bibitem [{\citenamefont {Yang}\ \emph {et~al.}(2023{\natexlab{c}})\citenamefont {Yang}, \citenamefont {Zhang},\ and\ \citenamefont {Zhang}}]{yang2023minimal}%
  \BibitemOpen
  \bibfield  {author} {\bibinfo {author} {\bibfnamefont {Y.-f.}\ \bibnamefont {Yang}}, \bibinfo {author} {\bibfnamefont {G.-M.}\ \bibnamefont {Zhang}},\ and\ \bibinfo {author} {\bibfnamefont {F.-C.}\ \bibnamefont {Zhang}},\ }\bibfield  {title} {\bibinfo {title} {Minimal effective model and possible high-{$T_c$} mechanism for superconductivity of {La$_{3}$Ni$_{2}$O$_7$} under high pressure},\ }\href@noop {} {\bibfield  {journal} {\bibinfo  {journal} {arXiv preprint arXiv:2308.01176}\ } (\bibinfo {year} {2023}{\natexlab{c}})}\BibitemShut {NoStop}%
\bibitem [{\citenamefont {Wu}\ \emph {et~al.}(2023)\citenamefont {Wu}, \citenamefont {Luo}, \citenamefont {Yao},\ and\ \citenamefont {Wang}}]{wu2023charge}%
  \BibitemOpen
  \bibfield  {author} {\bibinfo {author} {\bibfnamefont {W.}~\bibnamefont {Wu}}, \bibinfo {author} {\bibfnamefont {Z.}~\bibnamefont {Luo}}, \bibinfo {author} {\bibfnamefont {D.-X.}\ \bibnamefont {Yao}},\ and\ \bibinfo {author} {\bibfnamefont {M.}~\bibnamefont {Wang}},\ }\href@noop {} {\bibinfo {title} {Charge transfer and zhang-rice singlet bands in the nickelate superconductor $\mathrm{La_3Ni_2O_7}$ under pressure}} (\bibinfo {year} {2023}),\ \Eprint {https://arxiv.org/abs/2307.05662} {arXiv:2307.05662 [cond-mat.str-el]} \BibitemShut {NoStop}%
\bibitem [{\citenamefont {Lu}\ \emph {et~al.}(2023{\natexlab{a}})\citenamefont {Lu}, \citenamefont {Pan}, \citenamefont {Yang},\ and\ \citenamefont {Wu}}]{lu2023interplay}%
  \BibitemOpen
  \bibfield  {author} {\bibinfo {author} {\bibfnamefont {C.}~\bibnamefont {Lu}}, \bibinfo {author} {\bibfnamefont {Z.}~\bibnamefont {Pan}}, \bibinfo {author} {\bibfnamefont {F.}~\bibnamefont {Yang}},\ and\ \bibinfo {author} {\bibfnamefont {C.}~\bibnamefont {Wu}},\ }\bibfield  {title} {\bibinfo {title} {Interplay of two $ e_g $ orbitals in superconducting under pressure},\ }\href@noop {} {\bibfield  {journal} {\bibinfo  {journal} {arXiv preprint arXiv:2310.02915}\ } (\bibinfo {year} {2023}{\natexlab{a}})}\BibitemShut {NoStop}%
\bibitem [{\citenamefont {Shilenko}\ and\ \citenamefont {Leonov}(2023)}]{shilenko2023correlated}%
  \BibitemOpen
  \bibfield  {author} {\bibinfo {author} {\bibfnamefont {D.~A.}\ \bibnamefont {Shilenko}}\ and\ \bibinfo {author} {\bibfnamefont {I.~V.}\ \bibnamefont {Leonov}},\ }\bibfield  {title} {\bibinfo {title} {Correlated electronic structure, orbital-selective behavior, and magnetic correlations in double-layer {${\mathrm{La}}_{3}{\mathrm{Ni}}_{2}{\mathrm{O}}_{7}$} under pressure},\ }\href {https://doi.org/10.1103/PhysRevB.108.125105} {\bibfield  {journal} {\bibinfo  {journal} {Phys. Rev. B}\ }\textbf {\bibinfo {volume} {108}},\ \bibinfo {pages} {125105} (\bibinfo {year} {2023})}\BibitemShut {NoStop}%
\bibitem [{\citenamefont {Zhang}\ \emph {et~al.}(2023{\natexlab{f}})\citenamefont {Zhang}, \citenamefont {Lin}, \citenamefont {Moreo},\ and\ \citenamefont {Dagotto}}]{zhang2023electronic}%
  \BibitemOpen
  \bibfield  {author} {\bibinfo {author} {\bibfnamefont {Y.}~\bibnamefont {Zhang}}, \bibinfo {author} {\bibfnamefont {L.-F.}\ \bibnamefont {Lin}}, \bibinfo {author} {\bibfnamefont {A.}~\bibnamefont {Moreo}},\ and\ \bibinfo {author} {\bibfnamefont {E.}~\bibnamefont {Dagotto}},\ }\bibfield  {title} {\bibinfo {title} {Electronic structure, dimer physics, orbital-selective behavior, and magnetic tendencies in the bilayer nickelate superconductor la 3 ni 2 o 7 under pressure},\ }\href@noop {} {\bibfield  {journal} {\bibinfo  {journal} {Physical Review B}\ }\textbf {\bibinfo {volume} {108}},\ \bibinfo {pages} {L180510} (\bibinfo {year} {2023}{\natexlab{f}})}\BibitemShut {NoStop}%
\bibitem [{\citenamefont {Geisler}\ \emph {et~al.}(2023)\citenamefont {Geisler}, \citenamefont {Hamlin}, \citenamefont {Stewart}, \citenamefont {Hennig},\ and\ \citenamefont {Hirschfeld}}]{geisler2023structural}%
  \BibitemOpen
  \bibfield  {author} {\bibinfo {author} {\bibfnamefont {B.}~\bibnamefont {Geisler}}, \bibinfo {author} {\bibfnamefont {J.~J.}\ \bibnamefont {Hamlin}}, \bibinfo {author} {\bibfnamefont {G.~R.}\ \bibnamefont {Stewart}}, \bibinfo {author} {\bibfnamefont {R.~G.}\ \bibnamefont {Hennig}},\ and\ \bibinfo {author} {\bibfnamefont {P.}~\bibnamefont {Hirschfeld}},\ }\bibfield  {title} {\bibinfo {title} {Structural transitions, octahedral rotations, and electronic properties of {$A_3$Ni$_2$O$_7$} rare-earth nickelates under high pressure},\ }\href@noop {} {\bibfield  {journal} {\bibinfo  {journal} {arXiv preprint arXiv:2309.15078}\ } (\bibinfo {year} {2023})}\BibitemShut {NoStop}%
\bibitem [{\citenamefont {Lange}\ \emph {et~al.}(2023)\citenamefont {Lange}, \citenamefont {Homeier}, \citenamefont {Demler}, \citenamefont {Schollw{\"o}ck}, \citenamefont {Grusdt},\ and\ \citenamefont {Bohrdt}}]{lange2023feshbach}%
  \BibitemOpen
  \bibfield  {author} {\bibinfo {author} {\bibfnamefont {H.}~\bibnamefont {Lange}}, \bibinfo {author} {\bibfnamefont {L.}~\bibnamefont {Homeier}}, \bibinfo {author} {\bibfnamefont {E.}~\bibnamefont {Demler}}, \bibinfo {author} {\bibfnamefont {U.}~\bibnamefont {Schollw{\"o}ck}}, \bibinfo {author} {\bibfnamefont {F.}~\bibnamefont {Grusdt}},\ and\ \bibinfo {author} {\bibfnamefont {A.}~\bibnamefont {Bohrdt}},\ }\bibfield  {title} {\bibinfo {title} {Feshbach resonance in a strongly repulsive bilayer model: a possible scenario for bilayer nickelate superconductors},\ }\href@noop {} {\bibfield  {journal} {\bibinfo  {journal} {arXiv preprint arXiv:2309.15843}\ } (\bibinfo {year} {2023})}\BibitemShut {NoStop}%
\bibitem [{\citenamefont {LaBollita}\ \emph {et~al.}(2023)\citenamefont {LaBollita}, \citenamefont {Pardo}, \citenamefont {Norman},\ and\ \citenamefont {Botana}}]{labollita2023electronic}%
  \BibitemOpen
  \bibfield  {author} {\bibinfo {author} {\bibfnamefont {H.}~\bibnamefont {LaBollita}}, \bibinfo {author} {\bibfnamefont {V.}~\bibnamefont {Pardo}}, \bibinfo {author} {\bibfnamefont {M.~R.}\ \bibnamefont {Norman}},\ and\ \bibinfo {author} {\bibfnamefont {A.~S.}\ \bibnamefont {Botana}},\ }\bibfield  {title} {\bibinfo {title} {Electronic structure and magnetic properties of {La$_{3}$Ni$_{2}$O$_7$} under pressure},\ }\href@noop {} {\bibfield  {journal} {\bibinfo  {journal} {arXiv preprint arXiv:2309.17279}\ } (\bibinfo {year} {2023})}\BibitemShut {NoStop}%
\bibitem [{\citenamefont {Rhodes}\ and\ \citenamefont {Wahl}(2023)}]{rhodes2023structural}%
  \BibitemOpen
  \bibfield  {author} {\bibinfo {author} {\bibfnamefont {L.~C.}\ \bibnamefont {Rhodes}}\ and\ \bibinfo {author} {\bibfnamefont {P.}~\bibnamefont {Wahl}},\ }\bibfield  {title} {\bibinfo {title} {Structural routes to stabilise superconducting {{La$_{3}$Ni$_{2}$O$_{7}$}} at ambient pressure},\ }\href@noop {} {\bibfield  {journal} {\bibinfo  {journal} {arXiv preprint arXiv:2309.15745}\ } (\bibinfo {year} {2023})}\BibitemShut {NoStop}%
\bibitem [{\citenamefont {Zhang}\ \emph {et~al.}(2023{\natexlab{g}})\citenamefont {Zhang}, \citenamefont {Zhang}, \citenamefont {You},\ and\ \citenamefont {Weng}}]{zhang2023strong}%
  \BibitemOpen
  \bibfield  {author} {\bibinfo {author} {\bibfnamefont {J.-X.}\ \bibnamefont {Zhang}}, \bibinfo {author} {\bibfnamefont {H.-K.}\ \bibnamefont {Zhang}}, \bibinfo {author} {\bibfnamefont {Y.-Z.}\ \bibnamefont {You}},\ and\ \bibinfo {author} {\bibfnamefont {Z.-Y.}\ \bibnamefont {Weng}},\ }\bibfield  {title} {\bibinfo {title} {Strong pairing originated from an emergent $\mathbb{Z}_2$ berry phase in {La$_3$Ni$_2$O$_7$}},\ }\href@noop {} {\bibfield  {journal} {\bibinfo  {journal} {arXiv preprint arXiv:2309.05726}\ } (\bibinfo {year} {2023}{\natexlab{g}})},\ \Eprint {https://arxiv.org/abs/2309.05726} {arXiv:2309.05726 [cond-mat.str-el]} \BibitemShut {NoStop}%
\bibitem [{\citenamefont {Lu}\ \emph {et~al.}(2023{\natexlab{b}})\citenamefont {Lu}, \citenamefont {Pan}, \citenamefont {Yang},\ and\ \citenamefont {Wu}}]{lu2023interlayer}%
  \BibitemOpen
  \bibfield  {author} {\bibinfo {author} {\bibfnamefont {C.}~\bibnamefont {Lu}}, \bibinfo {author} {\bibfnamefont {Z.}~\bibnamefont {Pan}}, \bibinfo {author} {\bibfnamefont {F.}~\bibnamefont {Yang}},\ and\ \bibinfo {author} {\bibfnamefont {C.}~\bibnamefont {Wu}},\ }\bibfield  {title} {\bibinfo {title} {Interlayer coupling driven high-temperature superconductivity in {{La$_{3}$Ni$_{2}$O$_{7}$}} under pressure},\ }\href@noop {} {\bibfield  {journal} {\bibinfo  {journal} {arXiv preprint arXiv:2307.14965}\ } (\bibinfo {year} {2023}{\natexlab{b}})}\BibitemShut {NoStop}%
\bibitem [{\citenamefont {Chen}\ \emph {et~al.}(2023{\natexlab{b}})\citenamefont {Chen}, \citenamefont {Jiang}, \citenamefont {Li}, \citenamefont {Zhong},\ and\ \citenamefont {Lu}}]{chen2023critical}%
  \BibitemOpen
  \bibfield  {author} {\bibinfo {author} {\bibfnamefont {X.}~\bibnamefont {Chen}}, \bibinfo {author} {\bibfnamefont {P.}~\bibnamefont {Jiang}}, \bibinfo {author} {\bibfnamefont {J.}~\bibnamefont {Li}}, \bibinfo {author} {\bibfnamefont {Z.}~\bibnamefont {Zhong}},\ and\ \bibinfo {author} {\bibfnamefont {Y.}~\bibnamefont {Lu}},\ }\bibfield  {title} {\bibinfo {title} {Critical charge and spin instabilities in superconducting {{La$_{3}$Ni$_{2}$O$_{7}$}}},\ }\href@noop {} {\bibfield  {journal} {\bibinfo  {journal} {arXiv preprint arXiv:2307.07154}\ } (\bibinfo {year} {2023}{\natexlab{b}})}\BibitemShut {NoStop}%
\bibitem [{\citenamefont {Cao}\ and\ \citenamefont {Yang}(2023)}]{cao2023flat}%
  \BibitemOpen
  \bibfield  {author} {\bibinfo {author} {\bibfnamefont {Y.}~\bibnamefont {Cao}}\ and\ \bibinfo {author} {\bibfnamefont {Y.-f.}\ \bibnamefont {Yang}},\ }\bibfield  {title} {\bibinfo {title} {Flat bands promoted by hund's rule coupling in the candidate double-layer high-temperature superconductor {{La$_{3}$Ni$_{2}$O$_{7}$}}},\ }\href@noop {} {\bibfield  {journal} {\bibinfo  {journal} {arXiv preprint arXiv:2307.06806}\ } (\bibinfo {year} {2023})}\BibitemShut {NoStop}%
\bibitem [{\citenamefont {Christiansson}\ \emph {et~al.}(2023)\citenamefont {Christiansson}, \citenamefont {Petocchi},\ and\ \citenamefont {Werner}}]{christiansson2023correlated}%
  \BibitemOpen
  \bibfield  {author} {\bibinfo {author} {\bibfnamefont {V.}~\bibnamefont {Christiansson}}, \bibinfo {author} {\bibfnamefont {F.}~\bibnamefont {Petocchi}},\ and\ \bibinfo {author} {\bibfnamefont {P.}~\bibnamefont {Werner}},\ }\bibfield  {title} {\bibinfo {title} {Correlated electronic structure of {{La$_{3}$Ni$_{2}$O$_{7}$}} under pressure},\ }\href@noop {} {\bibfield  {journal} {\bibinfo  {journal} {arXiv preprint arXiv:2306.07931}\ } (\bibinfo {year} {2023})}\BibitemShut {NoStop}%
\bibitem [{\citenamefont {Zheng}\ and\ \citenamefont {Wú}(2023)}]{zheng2023superconductivity}%
  \BibitemOpen
  \bibfield  {author} {\bibinfo {author} {\bibfnamefont {Y.-Y.}\ \bibnamefont {Zheng}}\ and\ \bibinfo {author} {\bibfnamefont {W.}~\bibnamefont {Wú}},\ }\href@noop {} {\bibinfo {title} {Superconductivity in the bilayer two-orbital hubbard model}} (\bibinfo {year} {2023}),\ \Eprint {https://arxiv.org/abs/2312.03605} {arXiv:2312.03605 [cond-mat.str-el]} \BibitemShut {NoStop}%
\bibitem [{\citenamefont {Kakoi}\ \emph {et~al.}(2023)\citenamefont {Kakoi}, \citenamefont {Kaneko}, \citenamefont {Sakakibara}, \citenamefont {Ochi},\ and\ \citenamefont {Kuroki}}]{kakoi2023pair}%
  \BibitemOpen
  \bibfield  {author} {\bibinfo {author} {\bibfnamefont {M.}~\bibnamefont {Kakoi}}, \bibinfo {author} {\bibfnamefont {T.}~\bibnamefont {Kaneko}}, \bibinfo {author} {\bibfnamefont {H.}~\bibnamefont {Sakakibara}}, \bibinfo {author} {\bibfnamefont {M.}~\bibnamefont {Ochi}},\ and\ \bibinfo {author} {\bibfnamefont {K.}~\bibnamefont {Kuroki}},\ }\href@noop {} {\bibinfo {title} {Pair correlations of the hybridized orbitals in a ladder model for the bilayer nickelate {La$_{3}$Ni$_{2}$O$_7$}}} (\bibinfo {year} {2023}),\ \Eprint {https://arxiv.org/abs/2312.04304} {arXiv:2312.04304 [cond-mat.supr-con]} \BibitemShut {NoStop}%
\bibitem [{\citenamefont {Fan}\ \emph {et~al.}(2023)\citenamefont {Fan}, \citenamefont {Zhang}, \citenamefont {Zhan}, \citenamefont {Lv}, \citenamefont {Jiang}, \citenamefont {Normand},\ and\ \citenamefont {Xiang}}]{fan2023superconductivity}%
  \BibitemOpen
  \bibfield  {author} {\bibinfo {author} {\bibfnamefont {Z.}~\bibnamefont {Fan}}, \bibinfo {author} {\bibfnamefont {J.-F.}\ \bibnamefont {Zhang}}, \bibinfo {author} {\bibfnamefont {B.}~\bibnamefont {Zhan}}, \bibinfo {author} {\bibfnamefont {D.}~\bibnamefont {Lv}}, \bibinfo {author} {\bibfnamefont {X.-Y.}\ \bibnamefont {Jiang}}, \bibinfo {author} {\bibfnamefont {B.}~\bibnamefont {Normand}},\ and\ \bibinfo {author} {\bibfnamefont {T.}~\bibnamefont {Xiang}},\ }\bibfield  {title} {\bibinfo {title} {Superconductivity in nickelate and cuprate superconductors with strong bilayer coupling},\ }\href@noop {} {\bibfield  {journal} {\bibinfo  {journal} {arXiv preprint arXiv:2312.17064}\ } (\bibinfo {year} {2023})}\BibitemShut {NoStop}%
\bibitem [{\citenamefont {Geisler}\ \emph {et~al.}(2024)\citenamefont {Geisler}, \citenamefont {Fanfarillo}, \citenamefont {Hamlin}, \citenamefont {Stewart}, \citenamefont {Hennig},\ and\ \citenamefont {Hirschfeld}}]{geisler2024optical}%
  \BibitemOpen
  \bibfield  {author} {\bibinfo {author} {\bibfnamefont {B.}~\bibnamefont {Geisler}}, \bibinfo {author} {\bibfnamefont {L.}~\bibnamefont {Fanfarillo}}, \bibinfo {author} {\bibfnamefont {J.~J.}\ \bibnamefont {Hamlin}}, \bibinfo {author} {\bibfnamefont {G.~R.}\ \bibnamefont {Stewart}}, \bibinfo {author} {\bibfnamefont {R.~G.}\ \bibnamefont {Hennig}},\ and\ \bibinfo {author} {\bibfnamefont {P.~J.}\ \bibnamefont {Hirschfeld}},\ }\href@noop {} {\bibinfo {title} {Optical properties and electronic correlations in {La$_3$Ni$_2$O$_{7-\delta}$} bilayer nickelates under high pressure}} (\bibinfo {year} {2024}),\ \Eprint {https://arxiv.org/abs/2401.04258} {arXiv:2401.04258 [cond-mat.supr-con]} \BibitemShut {NoStop}%
\bibitem [{\citenamefont {Sakakibara}\ \emph {et~al.}(2024)\citenamefont {Sakakibara}, \citenamefont {Kitamine}, \citenamefont {Ochi},\ and\ \citenamefont {Kuroki}}]{sakakibara2024possible}%
  \BibitemOpen
  \bibfield  {author} {\bibinfo {author} {\bibfnamefont {H.}~\bibnamefont {Sakakibara}}, \bibinfo {author} {\bibfnamefont {N.}~\bibnamefont {Kitamine}}, \bibinfo {author} {\bibfnamefont {M.}~\bibnamefont {Ochi}},\ and\ \bibinfo {author} {\bibfnamefont {K.}~\bibnamefont {Kuroki}},\ }\href@noop {} {\bibinfo {title} {Possible high $t_c$ superconductivity in la$_3$ni$_2$o$_7$ under high pressure through manifestation of a nearly-half-filled bilayer hubbard model}} (\bibinfo {year} {2024}),\ \Eprint {https://arxiv.org/abs/2306.06039} {arXiv:2306.06039 [cond-mat.supr-con]} \BibitemShut {NoStop}%
\bibitem [{\citenamefont {Puphal}\ \emph {et~al.}(2023)\citenamefont {Puphal}, \citenamefont {Reiss}, \citenamefont {Enderlein}, \citenamefont {Wu}, \citenamefont {Khaliullin}, \citenamefont {Sundaramurthy}, \citenamefont {Priessnitz}, \citenamefont {Knauft}, \citenamefont {Richter}, \citenamefont {Isobe}, \citenamefont {van Aken}, \citenamefont {Takagi}, \citenamefont {Keimer}, \citenamefont {Suyolcu}, \citenamefont {Wehinger}, \citenamefont {Hansmann},\ and\ \citenamefont {Hepting}}]{puphal2023unconventional}%
  \BibitemOpen
  \bibfield  {author} {\bibinfo {author} {\bibfnamefont {P.}~\bibnamefont {Puphal}}, \bibinfo {author} {\bibfnamefont {P.}~\bibnamefont {Reiss}}, \bibinfo {author} {\bibfnamefont {N.}~\bibnamefont {Enderlein}}, \bibinfo {author} {\bibfnamefont {Y.-M.}\ \bibnamefont {Wu}}, \bibinfo {author} {\bibfnamefont {G.}~\bibnamefont {Khaliullin}}, \bibinfo {author} {\bibfnamefont {V.}~\bibnamefont {Sundaramurthy}}, \bibinfo {author} {\bibfnamefont {T.}~\bibnamefont {Priessnitz}}, \bibinfo {author} {\bibfnamefont {M.}~\bibnamefont {Knauft}}, \bibinfo {author} {\bibfnamefont {L.}~\bibnamefont {Richter}}, \bibinfo {author} {\bibfnamefont {M.}~\bibnamefont {Isobe}}, \bibinfo {author} {\bibfnamefont {P.~A.}\ \bibnamefont {van Aken}}, \bibinfo {author} {\bibfnamefont {H.}~\bibnamefont {Takagi}}, \bibinfo {author} {\bibfnamefont {B.}~\bibnamefont {Keimer}}, \bibinfo {author} {\bibfnamefont {Y.~E.}\ \bibnamefont {Suyolcu}}, \bibinfo {author} {\bibfnamefont {B.}~\bibnamefont {Wehinger}}, \bibinfo {author} {\bibfnamefont
  {P.}~\bibnamefont {Hansmann}},\ and\ \bibinfo {author} {\bibfnamefont {M.}~\bibnamefont {Hepting}},\ }\href@noop {} {\bibinfo {title} {Unconventional crystal structure of the high-pressure superconductor {La$_{3}$Ni$_{2}$O$_7$}}} (\bibinfo {year} {2023}),\ \Eprint {https://arxiv.org/abs/2312.07341} {arXiv:2312.07341 [cond-mat.supr-con]} \BibitemShut {NoStop}%
\bibitem [{\citenamefont {Chen}\ \emph {et~al.}(2023{\natexlab{c}})\citenamefont {Chen}, \citenamefont {Zhang}, \citenamefont {Thind}, \citenamefont {Sharma}, \citenamefont {LaBollita}, \citenamefont {Peterson}, \citenamefont {Zheng}, \citenamefont {Phelan}, \citenamefont {Botana}, \citenamefont {Klie},\ and\ \citenamefont {Mitchell}}]{chen2023polymorphism}%
  \BibitemOpen
  \bibfield  {author} {\bibinfo {author} {\bibfnamefont {X.}~\bibnamefont {Chen}}, \bibinfo {author} {\bibfnamefont {J.}~\bibnamefont {Zhang}}, \bibinfo {author} {\bibfnamefont {A.~S.}\ \bibnamefont {Thind}}, \bibinfo {author} {\bibfnamefont {S.}~\bibnamefont {Sharma}}, \bibinfo {author} {\bibfnamefont {H.}~\bibnamefont {LaBollita}}, \bibinfo {author} {\bibfnamefont {G.}~\bibnamefont {Peterson}}, \bibinfo {author} {\bibfnamefont {H.}~\bibnamefont {Zheng}}, \bibinfo {author} {\bibfnamefont {D.}~\bibnamefont {Phelan}}, \bibinfo {author} {\bibfnamefont {A.~S.}\ \bibnamefont {Botana}}, \bibinfo {author} {\bibfnamefont {R.~F.}\ \bibnamefont {Klie}},\ and\ \bibinfo {author} {\bibfnamefont {J.~F.}\ \bibnamefont {Mitchell}},\ }\href@noop {} {\bibinfo {title} {Polymorphism in ruddlesden-popper {La$_{3}$Ni$_{2}$O$_7$}: Discovery of a hidden phase with distinctive layer stacking}} (\bibinfo {year} {2023}{\natexlab{c}}),\ \Eprint {https://arxiv.org/abs/2312.06081} {arXiv:2312.06081 [cond-mat.supr-con]} \BibitemShut
  {NoStop}%
\bibitem [{\citenamefont {Dagotto}\ \emph {et~al.}(1992)\citenamefont {Dagotto}, \citenamefont {Riera},\ and\ \citenamefont {Scalapino}}]{dagotto1992superconductivity}%
  \BibitemOpen
  \bibfield  {author} {\bibinfo {author} {\bibfnamefont {E.}~\bibnamefont {Dagotto}}, \bibinfo {author} {\bibfnamefont {J.}~\bibnamefont {Riera}},\ and\ \bibinfo {author} {\bibfnamefont {D.}~\bibnamefont {Scalapino}},\ }\bibfield  {title} {\bibinfo {title} {Superconductivity in ladders and coupled planes},\ }\href@noop {} {\bibfield  {journal} {\bibinfo  {journal} {Physical Review B}\ }\textbf {\bibinfo {volume} {45}},\ \bibinfo {pages} {5744} (\bibinfo {year} {1992})}\BibitemShut {NoStop}%
\bibitem [{\citenamefont {Chen}\ \emph {et~al.}(2023{\natexlab{d}})\citenamefont {Chen}, \citenamefont {Liu}, \citenamefont {Jiao}, \citenamefont {Zou}, \citenamefont {Luo}, \citenamefont {Wu}, \citenamefont {Zhang}, \citenamefont {Guo},\ and\ \citenamefont {Shu}}]{chen2023evidence}%
  \BibitemOpen
  \bibfield  {author} {\bibinfo {author} {\bibfnamefont {K.}~\bibnamefont {Chen}}, \bibinfo {author} {\bibfnamefont {X.}~\bibnamefont {Liu}}, \bibinfo {author} {\bibfnamefont {J.}~\bibnamefont {Jiao}}, \bibinfo {author} {\bibfnamefont {M.}~\bibnamefont {Zou}}, \bibinfo {author} {\bibfnamefont {Y.}~\bibnamefont {Luo}}, \bibinfo {author} {\bibfnamefont {Q.}~\bibnamefont {Wu}}, \bibinfo {author} {\bibfnamefont {N.}~\bibnamefont {Zhang}}, \bibinfo {author} {\bibfnamefont {Y.}~\bibnamefont {Guo}},\ and\ \bibinfo {author} {\bibfnamefont {L.}~\bibnamefont {Shu}},\ }\href@noop {} {\bibinfo {title} {Evidence of spin density waves in {La$_3$Ni$_2$O$_{7-\delta}$}}} (\bibinfo {year} {2023}{\natexlab{d}}),\ \Eprint {https://arxiv.org/abs/2311.15717} {arXiv:2311.15717 [cond-mat.str-el]} \BibitemShut {NoStop}%
\bibitem [{\citenamefont {Liu}\ \emph {et~al.}(2023{\natexlab{b}})\citenamefont {Liu}, \citenamefont {Sun}, \citenamefont {Huo}, \citenamefont {Ma}, \citenamefont {Ji}, \citenamefont {Yi}, \citenamefont {Li}, \citenamefont {Liu}, \citenamefont {Yu}, \citenamefont {Zhang} \emph {et~al.}}]{liu2023evidence}%
  \BibitemOpen
  \bibfield  {author} {\bibinfo {author} {\bibfnamefont {Z.}~\bibnamefont {Liu}}, \bibinfo {author} {\bibfnamefont {H.}~\bibnamefont {Sun}}, \bibinfo {author} {\bibfnamefont {M.}~\bibnamefont {Huo}}, \bibinfo {author} {\bibfnamefont {X.}~\bibnamefont {Ma}}, \bibinfo {author} {\bibfnamefont {Y.}~\bibnamefont {Ji}}, \bibinfo {author} {\bibfnamefont {E.}~\bibnamefont {Yi}}, \bibinfo {author} {\bibfnamefont {L.}~\bibnamefont {Li}}, \bibinfo {author} {\bibfnamefont {H.}~\bibnamefont {Liu}}, \bibinfo {author} {\bibfnamefont {J.}~\bibnamefont {Yu}}, \bibinfo {author} {\bibfnamefont {Z.}~\bibnamefont {Zhang}}, \emph {et~al.},\ }\bibfield  {title} {\bibinfo {title} {Evidence for charge and spin density waves in single crystals of {La$_{3}$Ni$_{2}$O$_7$} and {La$_3$Ni$_2$O$_6$}},\ }\href@noop {} {\bibfield  {journal} {\bibinfo  {journal} {Science China Physics, Mechanics \& Astronomy}\ }\textbf {\bibinfo {volume} {66}},\ \bibinfo {pages} {217411} (\bibinfo {year} {2023}{\natexlab{b}})}\BibitemShut {NoStop}%
\bibitem [{\citenamefont {Yu}\ \emph {et~al.}(2009)\citenamefont {Yu}, \citenamefont {Li}, \citenamefont {Motoyama},\ and\ \citenamefont {Greven}}]{yu2009universal}%
  \BibitemOpen
  \bibfield  {author} {\bibinfo {author} {\bibfnamefont {G.}~\bibnamefont {Yu}}, \bibinfo {author} {\bibfnamefont {Y.}~\bibnamefont {Li}}, \bibinfo {author} {\bibfnamefont {E.}~\bibnamefont {Motoyama}},\ and\ \bibinfo {author} {\bibfnamefont {M.}~\bibnamefont {Greven}},\ }\bibfield  {title} {\bibinfo {title} {A universal relationship between magnetic resonance and superconducting gap in unconventional superconductors},\ }\href@noop {} {\bibfield  {journal} {\bibinfo  {journal} {Nature Physics}\ }\textbf {\bibinfo {volume} {5}},\ \bibinfo {pages} {873} (\bibinfo {year} {2009})}\BibitemShut {NoStop}%
\bibitem [{\citenamefont {Rossat-Mignod}\ \emph {et~al.}(1991)\citenamefont {Rossat-Mignod}, \citenamefont {Regnault}, \citenamefont {Vettier}, \citenamefont {Bourges}, \citenamefont {Burlet}, \citenamefont {Bossy}, \citenamefont {Henry},\ and\ \citenamefont {Lapertot}}]{rossat1991neutron}%
  \BibitemOpen
  \bibfield  {author} {\bibinfo {author} {\bibfnamefont {J.}~\bibnamefont {Rossat-Mignod}}, \bibinfo {author} {\bibfnamefont {L.}~\bibnamefont {Regnault}}, \bibinfo {author} {\bibfnamefont {C.}~\bibnamefont {Vettier}}, \bibinfo {author} {\bibfnamefont {P.}~\bibnamefont {Bourges}}, \bibinfo {author} {\bibfnamefont {P.}~\bibnamefont {Burlet}}, \bibinfo {author} {\bibfnamefont {J.}~\bibnamefont {Bossy}}, \bibinfo {author} {\bibfnamefont {J.}~\bibnamefont {Henry}},\ and\ \bibinfo {author} {\bibfnamefont {G.}~\bibnamefont {Lapertot}},\ }\bibfield  {title} {\bibinfo {title} {Neutron scattering study of the {YBa$_2$Cu$_3$O$_{6+x}$} system},\ }\href@noop {} {\bibfield  {journal} {\bibinfo  {journal} {Physica C: Superconductivity}\ }\textbf {\bibinfo {volume} {185}},\ \bibinfo {pages} {86} (\bibinfo {year} {1991})}\BibitemShut {NoStop}%
\bibitem [{\citenamefont {Mook}\ \emph {et~al.}(1993)\citenamefont {Mook}, \citenamefont {Yethiraj}, \citenamefont {Aeppli}, \citenamefont {Mason},\ and\ \citenamefont {Armstrong}}]{mookPolarized}%
  \BibitemOpen
  \bibfield  {author} {\bibinfo {author} {\bibfnamefont {H.~A.}\ \bibnamefont {Mook}}, \bibinfo {author} {\bibfnamefont {M.}~\bibnamefont {Yethiraj}}, \bibinfo {author} {\bibfnamefont {G.}~\bibnamefont {Aeppli}}, \bibinfo {author} {\bibfnamefont {T.~E.}\ \bibnamefont {Mason}},\ and\ \bibinfo {author} {\bibfnamefont {T.}~\bibnamefont {Armstrong}},\ }\bibfield  {title} {\bibinfo {title} {Polarized neutron determination of the magnetic excitations in {${\mathrm{YBa}}_{2}$${\mathrm{Cu}}_{3}$${\mathrm{O}}_{7}$}},\ }\href {https://doi.org/10.1103/PhysRevLett.70.3490} {\bibfield  {journal} {\bibinfo  {journal} {Phys. Rev. Lett.}\ }\textbf {\bibinfo {volume} {70}},\ \bibinfo {pages} {3490} (\bibinfo {year} {1993})}\BibitemShut {NoStop}%
\bibitem [{\citenamefont {Bourges}\ \emph {et~al.}(1996)\citenamefont {Bourges}, \citenamefont {Regnault}, \citenamefont {Sidis},\ and\ \citenamefont {Vettier}}]{Bourges1996}%
  \BibitemOpen
  \bibfield  {author} {\bibinfo {author} {\bibfnamefont {P.}~\bibnamefont {Bourges}}, \bibinfo {author} {\bibfnamefont {L.~P.}\ \bibnamefont {Regnault}}, \bibinfo {author} {\bibfnamefont {Y.}~\bibnamefont {Sidis}},\ and\ \bibinfo {author} {\bibfnamefont {C.}~\bibnamefont {Vettier}},\ }\bibfield  {title} {\bibinfo {title} {Inelastic-neutron-scattering study of antiferromagnetic fluctuations in {${\mathrm{YBa}}_{2}$${\mathrm{Cu}}_{3}$${\mathrm{O}}_{6.97}$}},\ }\href {https://doi.org/10.1103/PhysRevB.53.876} {\bibfield  {journal} {\bibinfo  {journal} {Phys. Rev. B}\ }\textbf {\bibinfo {volume} {53}},\ \bibinfo {pages} {876} (\bibinfo {year} {1996})}\BibitemShut {NoStop}%
\bibitem [{\citenamefont {Dai}\ \emph {et~al.}(1999)\citenamefont {Dai}, \citenamefont {Mook}, \citenamefont {Hayden}, \citenamefont {Aeppli}, \citenamefont {Perring}, \citenamefont {Hunt},\ and\ \citenamefont {Do$\check{\text{g}}$an}}]{Dai1999}%
  \BibitemOpen
  \bibfield  {author} {\bibinfo {author} {\bibfnamefont {P.}~\bibnamefont {Dai}}, \bibinfo {author} {\bibfnamefont {H.~A.}\ \bibnamefont {Mook}}, \bibinfo {author} {\bibfnamefont {S.~M.}\ \bibnamefont {Hayden}}, \bibinfo {author} {\bibfnamefont {G.}~\bibnamefont {Aeppli}}, \bibinfo {author} {\bibfnamefont {T.~G.}\ \bibnamefont {Perring}}, \bibinfo {author} {\bibfnamefont {R.~D.}\ \bibnamefont {Hunt}},\ and\ \bibinfo {author} {\bibfnamefont {F.}~\bibnamefont {Do$\check{\text{g}}$an}},\ }\bibfield  {title} {\bibinfo {title} {The magnetic excitation spectrum and thermodynamics of high-{T$_c$} superconductors},\ }\href {https://doi.org/10.1126/science.284.5418.1344} {\bibfield  {journal} {\bibinfo  {journal} {Science}\ }\textbf {\bibinfo {volume} {284}},\ \bibinfo {pages} {1344–1347} (\bibinfo {year} {1999})}\BibitemShut {NoStop}%
\bibitem [{\citenamefont {Hinkov}\ \emph {et~al.}(2007)\citenamefont {Hinkov}, \citenamefont {Bourges}, \citenamefont {Pailhès}, \citenamefont {Sidis}, \citenamefont {Ivanov}, \citenamefont {Frost}, \citenamefont {Perring}, \citenamefont {Lin}, \citenamefont {Chen},\ and\ \citenamefont {Keimer}}]{Hinkov2007}%
  \BibitemOpen
  \bibfield  {author} {\bibinfo {author} {\bibfnamefont {V.}~\bibnamefont {Hinkov}}, \bibinfo {author} {\bibfnamefont {P.}~\bibnamefont {Bourges}}, \bibinfo {author} {\bibfnamefont {S.}~\bibnamefont {Pailhès}}, \bibinfo {author} {\bibfnamefont {Y.}~\bibnamefont {Sidis}}, \bibinfo {author} {\bibfnamefont {A.}~\bibnamefont {Ivanov}}, \bibinfo {author} {\bibfnamefont {C.~D.}\ \bibnamefont {Frost}}, \bibinfo {author} {\bibfnamefont {T.~G.}\ \bibnamefont {Perring}}, \bibinfo {author} {\bibfnamefont {C.~T.}\ \bibnamefont {Lin}}, \bibinfo {author} {\bibfnamefont {D.~P.}\ \bibnamefont {Chen}},\ and\ \bibinfo {author} {\bibfnamefont {B.}~\bibnamefont {Keimer}},\ }\bibfield  {title} {\bibinfo {title} {Spin dynamics in the pseudogap state of a high-temperature superconductor},\ }\href {https://doi.org/10.1038/nphys720} {\bibfield  {journal} {\bibinfo  {journal} {Nature Physics}\ }\textbf {\bibinfo {volume} {3}},\ \bibinfo {pages} {780–785} (\bibinfo {year} {2007})}\BibitemShut {NoStop}%
\bibitem [{\citenamefont {Christianson}\ \emph {et~al.}(2008)\citenamefont {Christianson}, \citenamefont {Goremychkin}, \citenamefont {Osborn}, \citenamefont {Rosenkranz}, \citenamefont {Lumsden}, \citenamefont {Malliakas}, \citenamefont {Todorov}, \citenamefont {Claus}, \citenamefont {Chung}, \citenamefont {Kanatzidis}, \citenamefont {Bewley},\ and\ \citenamefont {Guidi}}]{Christianson2008}%
  \BibitemOpen
  \bibfield  {author} {\bibinfo {author} {\bibfnamefont {A.~D.}\ \bibnamefont {Christianson}}, \bibinfo {author} {\bibfnamefont {E.~A.}\ \bibnamefont {Goremychkin}}, \bibinfo {author} {\bibfnamefont {R.}~\bibnamefont {Osborn}}, \bibinfo {author} {\bibfnamefont {S.}~\bibnamefont {Rosenkranz}}, \bibinfo {author} {\bibfnamefont {M.~D.}\ \bibnamefont {Lumsden}}, \bibinfo {author} {\bibfnamefont {C.~D.}\ \bibnamefont {Malliakas}}, \bibinfo {author} {\bibfnamefont {I.~S.}\ \bibnamefont {Todorov}}, \bibinfo {author} {\bibfnamefont {H.}~\bibnamefont {Claus}}, \bibinfo {author} {\bibfnamefont {D.~Y.}\ \bibnamefont {Chung}}, \bibinfo {author} {\bibfnamefont {M.~G.}\ \bibnamefont {Kanatzidis}}, \bibinfo {author} {\bibfnamefont {R.~I.}\ \bibnamefont {Bewley}},\ and\ \bibinfo {author} {\bibfnamefont {T.}~\bibnamefont {Guidi}},\ }\bibfield  {title} {\bibinfo {title} {Unconventional superconductivity in {Ba$_{0.6}$K$_{0.4}$Fe$_{2}$As$_{2}$} from inelastic neutron scattering},\ }\href {https://doi.org/10.1038/nature07625}
  {\bibfield  {journal} {\bibinfo  {journal} {Nature}\ }\textbf {\bibinfo {volume} {456}},\ \bibinfo {pages} {930–932} (\bibinfo {year} {2008})}\BibitemShut {NoStop}%
\bibitem [{\citenamefont {Inosov}\ \emph {et~al.}(2009)\citenamefont {Inosov}, \citenamefont {Park}, \citenamefont {Bourges}, \citenamefont {Sun}, \citenamefont {Sidis}, \citenamefont {Schneidewind}, \citenamefont {Hradil}, \citenamefont {Haug}, \citenamefont {Lin}, \citenamefont {Keimer},\ and\ \citenamefont {Hinkov}}]{Inosov2009}%
  \BibitemOpen
  \bibfield  {author} {\bibinfo {author} {\bibfnamefont {D.~S.}\ \bibnamefont {Inosov}}, \bibinfo {author} {\bibfnamefont {J.~T.}\ \bibnamefont {Park}}, \bibinfo {author} {\bibfnamefont {P.}~\bibnamefont {Bourges}}, \bibinfo {author} {\bibfnamefont {D.~L.}\ \bibnamefont {Sun}}, \bibinfo {author} {\bibfnamefont {Y.}~\bibnamefont {Sidis}}, \bibinfo {author} {\bibfnamefont {A.}~\bibnamefont {Schneidewind}}, \bibinfo {author} {\bibfnamefont {K.}~\bibnamefont {Hradil}}, \bibinfo {author} {\bibfnamefont {D.}~\bibnamefont {Haug}}, \bibinfo {author} {\bibfnamefont {C.~T.}\ \bibnamefont {Lin}}, \bibinfo {author} {\bibfnamefont {B.}~\bibnamefont {Keimer}},\ and\ \bibinfo {author} {\bibfnamefont {V.}~\bibnamefont {Hinkov}},\ }\bibfield  {title} {\bibinfo {title} {Normal-state spin dynamics and temperature-dependent spin-resonance energy in optimally doped {BaFe$_{1.85}$Co$_{0.15}$As$_2$}},\ }\href {https://doi.org/10.1038/nphys1483} {\bibfield  {journal} {\bibinfo  {journal} {Nature Physics}\ }\textbf {\bibinfo {volume}
  {6}},\ \bibinfo {pages} {178–181} (\bibinfo {year} {2009})}\BibitemShut {NoStop}%
\bibitem [{\citenamefont {Park}\ \emph {et~al.}(2011)\citenamefont {Park}, \citenamefont {Friemel}, \citenamefont {Li}, \citenamefont {Kim}, \citenamefont {Tsurkan}, \citenamefont {Deisenhofer}, \citenamefont {Krug~von Nidda}, \citenamefont {Loidl}, \citenamefont {Ivanov}, \citenamefont {Keimer},\ and\ \citenamefont {Inosov}}]{Park2011}%
  \BibitemOpen
  \bibfield  {author} {\bibinfo {author} {\bibfnamefont {J.~T.}\ \bibnamefont {Park}}, \bibinfo {author} {\bibfnamefont {G.}~\bibnamefont {Friemel}}, \bibinfo {author} {\bibfnamefont {Y.}~\bibnamefont {Li}}, \bibinfo {author} {\bibfnamefont {J.-H.}\ \bibnamefont {Kim}}, \bibinfo {author} {\bibfnamefont {V.}~\bibnamefont {Tsurkan}}, \bibinfo {author} {\bibfnamefont {J.}~\bibnamefont {Deisenhofer}}, \bibinfo {author} {\bibfnamefont {H.-A.}\ \bibnamefont {Krug~von Nidda}}, \bibinfo {author} {\bibfnamefont {A.}~\bibnamefont {Loidl}}, \bibinfo {author} {\bibfnamefont {A.}~\bibnamefont {Ivanov}}, \bibinfo {author} {\bibfnamefont {B.}~\bibnamefont {Keimer}},\ and\ \bibinfo {author} {\bibfnamefont {D.~S.}\ \bibnamefont {Inosov}},\ }\bibfield  {title} {\bibinfo {title} {Magnetic resonant mode in the low-energy spin-excitation spectrum of superconducting {${\mathrm{Rb}}_{2}{\mathrm{Fe}}_{4}{\mathrm{Se}}_{5}$} single crystals},\ }\href {https://doi.org/10.1103/PhysRevLett.107.177005} {\bibfield  {journal} {\bibinfo
  {journal} {Phys. Rev. Lett.}\ }\textbf {\bibinfo {volume} {107}},\ \bibinfo {pages} {177005} (\bibinfo {year} {2011})}\BibitemShut {NoStop}%
\bibitem [{\citenamefont {Dai}(2015)}]{Dai2015}%
  \BibitemOpen
  \bibfield  {author} {\bibinfo {author} {\bibfnamefont {P.}~\bibnamefont {Dai}},\ }\bibfield  {title} {\bibinfo {title} {Antiferromagnetic order and spin dynamics in iron-based superconductors},\ }\href {https://doi.org/10.1103/RevModPhys.87.855} {\bibfield  {journal} {\bibinfo  {journal} {Rev. Mod. Phys.}\ }\textbf {\bibinfo {volume} {87}},\ \bibinfo {pages} {855} (\bibinfo {year} {2015})}\BibitemShut {NoStop}%
\bibitem [{\citenamefont {Stock}\ \emph {et~al.}(2008)\citenamefont {Stock}, \citenamefont {Broholm}, \citenamefont {Hudis}, \citenamefont {Kang},\ and\ \citenamefont {Petrovic}}]{Stock2008}%
  \BibitemOpen
  \bibfield  {author} {\bibinfo {author} {\bibfnamefont {C.}~\bibnamefont {Stock}}, \bibinfo {author} {\bibfnamefont {C.}~\bibnamefont {Broholm}}, \bibinfo {author} {\bibfnamefont {J.}~\bibnamefont {Hudis}}, \bibinfo {author} {\bibfnamefont {H.~J.}\ \bibnamefont {Kang}},\ and\ \bibinfo {author} {\bibfnamefont {C.}~\bibnamefont {Petrovic}},\ }\bibfield  {title} {\bibinfo {title} {Spin resonance in the $d$-wave superconductor {${\mathrm{CeCoIn}}_{5}$}},\ }\href {https://doi.org/10.1103/PhysRevLett.100.087001} {\bibfield  {journal} {\bibinfo  {journal} {Phys. Rev. Lett.}\ }\textbf {\bibinfo {volume} {100}},\ \bibinfo {pages} {087001} (\bibinfo {year} {2008})}\BibitemShut {NoStop}%
\bibitem [{\citenamefont {Song}\ \emph {et~al.}(2020)\citenamefont {Song}, \citenamefont {Wang}, \citenamefont {S.~Van~Dyke}, \citenamefont {Pouse}, \citenamefont {Ran}, \citenamefont {Yazici}, \citenamefont {Schneidewind}, \citenamefont {Čermák}, \citenamefont {Qiu}, \citenamefont {Maple}, \citenamefont {Morr},\ and\ \citenamefont {Dai}}]{Song2020}%
  \BibitemOpen
  \bibfield  {author} {\bibinfo {author} {\bibfnamefont {Y.}~\bibnamefont {Song}}, \bibinfo {author} {\bibfnamefont {W.}~\bibnamefont {Wang}}, \bibinfo {author} {\bibfnamefont {J.}~\bibnamefont {S.~Van~Dyke}}, \bibinfo {author} {\bibfnamefont {N.}~\bibnamefont {Pouse}}, \bibinfo {author} {\bibfnamefont {S.}~\bibnamefont {Ran}}, \bibinfo {author} {\bibfnamefont {D.}~\bibnamefont {Yazici}}, \bibinfo {author} {\bibfnamefont {A.}~\bibnamefont {Schneidewind}}, \bibinfo {author} {\bibfnamefont {P.}~\bibnamefont {Čermák}}, \bibinfo {author} {\bibfnamefont {Y.}~\bibnamefont {Qiu}}, \bibinfo {author} {\bibfnamefont {M.~B.}\ \bibnamefont {Maple}}, \bibinfo {author} {\bibfnamefont {D.~K.}\ \bibnamefont {Morr}},\ and\ \bibinfo {author} {\bibfnamefont {P.}~\bibnamefont {Dai}},\ }\bibfield  {title} {\bibinfo {title} {Nature of the spin resonance mode in cecoin5},\ }\bibfield  {journal} {\bibinfo  {journal} {Communications Physics}\ }\textbf {\bibinfo {volume} {3}},\ \href {https://doi.org/10.1038/s42005-020-0365-8}
  {10.1038/s42005-020-0365-8} (\bibinfo {year} {2020})\BibitemShut {NoStop}%
\bibitem [{\citenamefont {Scalapino}(2012)}]{scalapino2012common}%
  \BibitemOpen
  \bibfield  {author} {\bibinfo {author} {\bibfnamefont {D.~J.}\ \bibnamefont {Scalapino}},\ }\bibfield  {title} {\bibinfo {title} {A common thread: The pairing interaction for unconventional superconductors},\ }\href@noop {} {\bibfield  {journal} {\bibinfo  {journal} {Reviews of Modern Physics}\ }\textbf {\bibinfo {volume} {84}},\ \bibinfo {pages} {1383} (\bibinfo {year} {2012})}\BibitemShut {NoStop}%
\bibitem [{\citenamefont {Bulut}\ \emph {et~al.}(1992)\citenamefont {Bulut}, \citenamefont {Scalapino},\ and\ \citenamefont {Scalettar}}]{bulut1992nodeless}%
  \BibitemOpen
  \bibfield  {author} {\bibinfo {author} {\bibfnamefont {N.}~\bibnamefont {Bulut}}, \bibinfo {author} {\bibfnamefont {D.~J.}\ \bibnamefont {Scalapino}},\ and\ \bibinfo {author} {\bibfnamefont {R.~T.}\ \bibnamefont {Scalettar}},\ }\bibfield  {title} {\bibinfo {title} {Nodeless d-wave pairing in a two-layer hubbard model},\ }\href@noop {} {\bibfield  {journal} {\bibinfo  {journal} {Physical Review B}\ }\textbf {\bibinfo {volume} {45}},\ \bibinfo {pages} {5577} (\bibinfo {year} {1992})}\BibitemShut {NoStop}%
\bibitem [{\citenamefont {Fong}\ \emph {et~al.}(1995)\citenamefont {Fong}, \citenamefont {Keimer}, \citenamefont {Anderson}, \citenamefont {Reznik}, \citenamefont {Do$\check{\text{g}}$an},\ and\ \citenamefont {Aksay}}]{Fong1996}%
  \BibitemOpen
  \bibfield  {author} {\bibinfo {author} {\bibfnamefont {H.~F.}\ \bibnamefont {Fong}}, \bibinfo {author} {\bibfnamefont {B.}~\bibnamefont {Keimer}}, \bibinfo {author} {\bibfnamefont {P.~W.}\ \bibnamefont {Anderson}}, \bibinfo {author} {\bibfnamefont {D.}~\bibnamefont {Reznik}}, \bibinfo {author} {\bibfnamefont {F.}~\bibnamefont {Do$\check{\text{g}}$an}},\ and\ \bibinfo {author} {\bibfnamefont {I.~A.}\ \bibnamefont {Aksay}},\ }\bibfield  {title} {\bibinfo {title} {Phonon and magnetic neutron scattering at 41 mev in {YB${\mathrm{a}}_{2}$C${\mathrm{u}}_{3}$${\mathrm{O}}_{7}$}},\ }\href {https://doi.org/10.1103/PhysRevLett.75.316} {\bibfield  {journal} {\bibinfo  {journal} {Phys. Rev. Lett.}\ }\textbf {\bibinfo {volume} {75}},\ \bibinfo {pages} {316} (\bibinfo {year} {1995})}\BibitemShut {NoStop}%
\bibitem [{\citenamefont {Abanov}\ and\ \citenamefont {Chubukov}(1999)}]{Abanov1999}%
  \BibitemOpen
  \bibfield  {author} {\bibinfo {author} {\bibfnamefont {A.}~\bibnamefont {Abanov}}\ and\ \bibinfo {author} {\bibfnamefont {A.~V.}\ \bibnamefont {Chubukov}},\ }\bibfield  {title} {\bibinfo {title} {A relation between the resonance neutron peak and arpes data in cuprates},\ }\href {https://doi.org/10.1103/PhysRevLett.83.1652} {\bibfield  {journal} {\bibinfo  {journal} {Phys. Rev. Lett.}\ }\textbf {\bibinfo {volume} {83}},\ \bibinfo {pages} {1652} (\bibinfo {year} {1999})}\BibitemShut {NoStop}%
\bibitem [{\citenamefont {Brinckmann}\ and\ \citenamefont {Lee}(1999)}]{Brinkmann1999}%
  \BibitemOpen
  \bibfield  {author} {\bibinfo {author} {\bibfnamefont {J.}~\bibnamefont {Brinckmann}}\ and\ \bibinfo {author} {\bibfnamefont {P.~A.}\ \bibnamefont {Lee}},\ }\bibfield  {title} {\bibinfo {title} {Slave boson approach to neutron scattering in ${{\mathrm{YBa}}_{2}{\mathrm{Cu}}_{3}O}_{6+\mathit{y}}$ superconductors},\ }\href {https://doi.org/10.1103/PhysRevLett.82.2915} {\bibfield  {journal} {\bibinfo  {journal} {Phys. Rev. Lett.}\ }\textbf {\bibinfo {volume} {82}},\ \bibinfo {pages} {2915} (\bibinfo {year} {1999})}\BibitemShut {NoStop}%
\bibitem [{\citenamefont {Norman}(2000)}]{Norman2000}%
  \BibitemOpen
  \bibfield  {author} {\bibinfo {author} {\bibfnamefont {M.~R.}\ \bibnamefont {Norman}},\ }\bibfield  {title} {\bibinfo {title} {Relation of neutron incommensurability to electronic structure in high-temperature superconductors},\ }\href {https://doi.org/10.1103/PhysRevB.61.14751} {\bibfield  {journal} {\bibinfo  {journal} {Phys. Rev. B}\ }\textbf {\bibinfo {volume} {61}},\ \bibinfo {pages} {14751} (\bibinfo {year} {2000})}\BibitemShut {NoStop}%
\bibitem [{\citenamefont {Kao}\ \emph {et~al.}(2000)\citenamefont {Kao}, \citenamefont {Si},\ and\ \citenamefont {Levin}}]{Kao2000}%
  \BibitemOpen
  \bibfield  {author} {\bibinfo {author} {\bibfnamefont {Y.-J.}\ \bibnamefont {Kao}}, \bibinfo {author} {\bibfnamefont {Q.}~\bibnamefont {Si}},\ and\ \bibinfo {author} {\bibfnamefont {K.}~\bibnamefont {Levin}},\ }\bibfield  {title} {\bibinfo {title} {Frequency evolution of neutron peaks below ${T}_{c}:$ commensurate and incommensurate structure in {${\mathrm{La}}_{0.85}{\mathrm{Sr}}_{0.15}{\mathrm{CuO}}_{4}$} and {${\mathrm{YBa}}_{2}{\mathrm{Cu}}_{3}{\mathrm{O}}_{6.6}$}},\ }\href {https://doi.org/10.1103/PhysRevB.61.R11898} {\bibfield  {journal} {\bibinfo  {journal} {Phys. Rev. B}\ }\textbf {\bibinfo {volume} {61}},\ \bibinfo {pages} {R11898} (\bibinfo {year} {2000})}\BibitemShut {NoStop}%
\bibitem [{\citenamefont {Chubukov}\ \emph {et~al.}(2001)\citenamefont {Chubukov}, \citenamefont {Jank\'o},\ and\ \citenamefont {Tchernyshyov}}]{Chubukov2001}%
  \BibitemOpen
  \bibfield  {author} {\bibinfo {author} {\bibfnamefont {A.~V.}\ \bibnamefont {Chubukov}}, \bibinfo {author} {\bibfnamefont {B.}~\bibnamefont {Jank\'o}},\ and\ \bibinfo {author} {\bibfnamefont {O.}~\bibnamefont {Tchernyshyov}},\ }\bibfield  {title} {\bibinfo {title} {Dispersion of the neutron resonance in cuprate superconductors},\ }\href {https://doi.org/10.1103/PhysRevB.63.180507} {\bibfield  {journal} {\bibinfo  {journal} {Phys. Rev. B}\ }\textbf {\bibinfo {volume} {63}},\ \bibinfo {pages} {180507} (\bibinfo {year} {2001})}\BibitemShut {NoStop}%
\bibitem [{\citenamefont {Onufrieva}\ and\ \citenamefont {Pfeuty}(2002)}]{Onufrieva2002}%
  \BibitemOpen
  \bibfield  {author} {\bibinfo {author} {\bibfnamefont {F.}~\bibnamefont {Onufrieva}}\ and\ \bibinfo {author} {\bibfnamefont {P.}~\bibnamefont {Pfeuty}},\ }\bibfield  {title} {\bibinfo {title} {Spin dynamics of a two-dimensional metal in a superconducting state: Application to the high-{T$_{c}$} cuprates},\ }\href {https://doi.org/10.1103/PhysRevB.65.054515} {\bibfield  {journal} {\bibinfo  {journal} {Phys. Rev. B}\ }\textbf {\bibinfo {volume} {65}},\ \bibinfo {pages} {054515} (\bibinfo {year} {2002})}\BibitemShut {NoStop}%
\bibitem [{\citenamefont {Eremin}\ \emph {et~al.}(2005)\citenamefont {Eremin}, \citenamefont {Morr}, \citenamefont {Chubukov}, \citenamefont {Bennemann},\ and\ \citenamefont {Norman}}]{Eremin2005}%
  \BibitemOpen
  \bibfield  {author} {\bibinfo {author} {\bibfnamefont {I.}~\bibnamefont {Eremin}}, \bibinfo {author} {\bibfnamefont {D.~K.}\ \bibnamefont {Morr}}, \bibinfo {author} {\bibfnamefont {A.~V.}\ \bibnamefont {Chubukov}}, \bibinfo {author} {\bibfnamefont {K.~H.}\ \bibnamefont {Bennemann}},\ and\ \bibinfo {author} {\bibfnamefont {M.~R.}\ \bibnamefont {Norman}},\ }\bibfield  {title} {\bibinfo {title} {Novel neutron resonance mode in ${d}_{{x}^{2}\ensuremath{-}{y}^{2}}$-wave superconductors},\ }\href {https://doi.org/10.1103/PhysRevLett.94.147001} {\bibfield  {journal} {\bibinfo  {journal} {Phys. Rev. Lett.}\ }\textbf {\bibinfo {volume} {94}},\ \bibinfo {pages} {147001} (\bibinfo {year} {2005})}\BibitemShut {NoStop}%
\bibitem [{\citenamefont {Sato}\ \emph {et~al.}(1988)\citenamefont {Sato}, \citenamefont {Shamoto}, \citenamefont {Tranquada}, \citenamefont {Shirane},\ and\ \citenamefont {Keimer}}]{sato1988two}%
  \BibitemOpen
  \bibfield  {author} {\bibinfo {author} {\bibfnamefont {M.}~\bibnamefont {Sato}}, \bibinfo {author} {\bibfnamefont {S.}~\bibnamefont {Shamoto}}, \bibinfo {author} {\bibfnamefont {J.}~\bibnamefont {Tranquada}}, \bibinfo {author} {\bibfnamefont {G.}~\bibnamefont {Shirane}},\ and\ \bibinfo {author} {\bibfnamefont {B.}~\bibnamefont {Keimer}},\ }\bibfield  {title} {\bibinfo {title} {Two-dimensional antiferromagnetic excitations from a large single crystal of {YBa$_2$Cu$_3$O$_{6.2}$}},\ }\href@noop {} {\bibfield  {journal} {\bibinfo  {journal} {Physical review letters}\ }\textbf {\bibinfo {volume} {61}},\ \bibinfo {pages} {1317} (\bibinfo {year} {1988})}\BibitemShut {NoStop}%
\bibitem [{\citenamefont {Blumberg}\ \emph {et~al.}(1995)\citenamefont {Blumberg}, \citenamefont {Stojkovi{\'c}},\ and\ \citenamefont {Klein}}]{blumberg1995antiferromagnetic}%
  \BibitemOpen
  \bibfield  {author} {\bibinfo {author} {\bibfnamefont {G.}~\bibnamefont {Blumberg}}, \bibinfo {author} {\bibfnamefont {B.~P.}\ \bibnamefont {Stojkovi{\'c}}},\ and\ \bibinfo {author} {\bibfnamefont {M.}~\bibnamefont {Klein}},\ }\bibfield  {title} {\bibinfo {title} {Antiferromagnetic excitations and van hove singularities in {YBa$_2$Cu$_3$O$_{6+x}$}},\ }\href@noop {} {\bibfield  {journal} {\bibinfo  {journal} {Physical Review B}\ }\textbf {\bibinfo {volume} {52}},\ \bibinfo {pages} {R15741} (\bibinfo {year} {1995})}\BibitemShut {NoStop}%
\bibitem [{\citenamefont {Normand}\ \emph {et~al.}(1995)\citenamefont {Normand}, \citenamefont {Kohno},\ and\ \citenamefont {Fukuyama}}]{Normand}%
  \BibitemOpen
  \bibfield  {author} {\bibinfo {author} {\bibfnamefont {B.}~\bibnamefont {Normand}}, \bibinfo {author} {\bibfnamefont {H.}~\bibnamefont {Kohno}},\ and\ \bibinfo {author} {\bibfnamefont {H.}~\bibnamefont {Fukuyama}},\ }\bibfield  {title} {\bibinfo {title} {Dynamic susceptibility and phonon anomalies in the bilayer t-j model},\ }\href {https://doi.org/10.1143/JPSJ.64.3903} {\bibfield  {journal} {\bibinfo  {journal} {Journal of the Physical Society of Japan}\ }\textbf {\bibinfo {volume} {64}},\ \bibinfo {pages} {3903} (\bibinfo {year} {1995})}\BibitemShut {NoStop}%
\bibitem [{\citenamefont {Eremin}\ \emph {et~al.}(2007)\citenamefont {Eremin}, \citenamefont {Morr}, \citenamefont {Chubukov},\ and\ \citenamefont {Bennemann}}]{eremin2007spin}%
  \BibitemOpen
  \bibfield  {author} {\bibinfo {author} {\bibfnamefont {I.}~\bibnamefont {Eremin}}, \bibinfo {author} {\bibfnamefont {D.~K.}\ \bibnamefont {Morr}}, \bibinfo {author} {\bibfnamefont {A.~V.}\ \bibnamefont {Chubukov}},\ and\ \bibinfo {author} {\bibfnamefont {K.}~\bibnamefont {Bennemann}},\ }\bibfield  {title} {\bibinfo {title} {Spin susceptibility in bilayered cuprates: Resonant magnetic excitations},\ }\href@noop {} {\bibfield  {journal} {\bibinfo  {journal} {Physical Review B}\ }\textbf {\bibinfo {volume} {75}},\ \bibinfo {pages} {184534} (\bibinfo {year} {2007})}\BibitemShut {NoStop}%
\bibitem [{\citenamefont {Headings}\ \emph {et~al.}(2011)\citenamefont {Headings}, \citenamefont {Hayden}, \citenamefont {Kulda}, \citenamefont {Babu},\ and\ \citenamefont {Cardwell}}]{headings2011spin}%
  \BibitemOpen
  \bibfield  {author} {\bibinfo {author} {\bibfnamefont {N.}~\bibnamefont {Headings}}, \bibinfo {author} {\bibfnamefont {S.}~\bibnamefont {Hayden}}, \bibinfo {author} {\bibfnamefont {J.}~\bibnamefont {Kulda}}, \bibinfo {author} {\bibfnamefont {N.~H.}\ \bibnamefont {Babu}},\ and\ \bibinfo {author} {\bibfnamefont {D.}~\bibnamefont {Cardwell}},\ }\bibfield  {title} {\bibinfo {title} {Spin anisotropy of the magnetic excitations in the normal and superconducting states of optimally doped {YBa$_2$Cu$_3$O$_{6.9}$} studied by polarized neutron spectroscopy},\ }\href@noop {} {\bibfield  {journal} {\bibinfo  {journal} {Physical Review B}\ }\textbf {\bibinfo {volume} {84}},\ \bibinfo {pages} {104513} (\bibinfo {year} {2011})}\BibitemShut {NoStop}%
\bibitem [{\citenamefont {Pailh\`es}\ \emph {et~al.}(2006)\citenamefont {Pailh\`es}, \citenamefont {Ulrich}, \citenamefont {Fauqu\'e}, \citenamefont {Hinkov}, \citenamefont {Sidis}, \citenamefont {Ivanov}, \citenamefont {Lin}, \citenamefont {Keimer},\ and\ \citenamefont {Bourges}}]{Pailhes2006}%
  \BibitemOpen
  \bibfield  {author} {\bibinfo {author} {\bibfnamefont {S.}~\bibnamefont {Pailh\`es}}, \bibinfo {author} {\bibfnamefont {C.}~\bibnamefont {Ulrich}}, \bibinfo {author} {\bibfnamefont {B.}~\bibnamefont {Fauqu\'e}}, \bibinfo {author} {\bibfnamefont {V.}~\bibnamefont {Hinkov}}, \bibinfo {author} {\bibfnamefont {Y.}~\bibnamefont {Sidis}}, \bibinfo {author} {\bibfnamefont {A.}~\bibnamefont {Ivanov}}, \bibinfo {author} {\bibfnamefont {C.~T.}\ \bibnamefont {Lin}}, \bibinfo {author} {\bibfnamefont {B.}~\bibnamefont {Keimer}},\ and\ \bibinfo {author} {\bibfnamefont {P.}~\bibnamefont {Bourges}},\ }\bibfield  {title} {\bibinfo {title} {Doping dependence of bilayer resonant spin excitations in $(\mathrm{Y},\mathrm{Ca}){\mathrm{ba}}_{2}{\mathrm{cu}}_{3}{\mathrm{o}}_{6+x}$},\ }\href {https://doi.org/10.1103/PhysRevLett.96.257001} {\bibfield  {journal} {\bibinfo  {journal} {Phys. Rev. Lett.}\ }\textbf {\bibinfo {volume} {96}},\ \bibinfo {pages} {257001} (\bibinfo {year} {2006})}\BibitemShut {NoStop}%
\bibitem [{\citenamefont {Capogna}\ \emph {et~al.}(2007)\citenamefont {Capogna}, \citenamefont {Fauqu\'e}, \citenamefont {Sidis}, \citenamefont {Ulrich}, \citenamefont {Bourges}, \citenamefont {Pailh\`es}, \citenamefont {Ivanov}, \citenamefont {Tallon}, \citenamefont {Liang}, \citenamefont {Lin}, \citenamefont {Rykov},\ and\ \citenamefont {Keimer}}]{Capogna2007}%
  \BibitemOpen
  \bibfield  {author} {\bibinfo {author} {\bibfnamefont {L.}~\bibnamefont {Capogna}}, \bibinfo {author} {\bibfnamefont {B.}~\bibnamefont {Fauqu\'e}}, \bibinfo {author} {\bibfnamefont {Y.}~\bibnamefont {Sidis}}, \bibinfo {author} {\bibfnamefont {C.}~\bibnamefont {Ulrich}}, \bibinfo {author} {\bibfnamefont {P.}~\bibnamefont {Bourges}}, \bibinfo {author} {\bibfnamefont {S.}~\bibnamefont {Pailh\`es}}, \bibinfo {author} {\bibfnamefont {A.}~\bibnamefont {Ivanov}}, \bibinfo {author} {\bibfnamefont {J.~L.}\ \bibnamefont {Tallon}}, \bibinfo {author} {\bibfnamefont {B.}~\bibnamefont {Liang}}, \bibinfo {author} {\bibfnamefont {C.~T.}\ \bibnamefont {Lin}}, \bibinfo {author} {\bibfnamefont {A.~I.}\ \bibnamefont {Rykov}},\ and\ \bibinfo {author} {\bibfnamefont {B.}~\bibnamefont {Keimer}},\ }\bibfield  {title} {\bibinfo {title} {Odd and even magnetic resonant modes in highly overdoped ${\mathrm{bi}}_{2}{\mathrm{sr}}_{2}\mathrm{Ca}{\mathrm{cu}}_{2}{\mathrm{o}}_{8+\ensuremath{\delta}}$},\ }\href
  {https://doi.org/10.1103/PhysRevB.75.060502} {\bibfield  {journal} {\bibinfo  {journal} {Phys. Rev. B}\ }\textbf {\bibinfo {volume} {75}},\ \bibinfo {pages} {060502} (\bibinfo {year} {2007})}\BibitemShut {NoStop}%
\bibitem [{\citenamefont {Graser}\ \emph {et~al.}(2009)\citenamefont {Graser}, \citenamefont {Maier}, \citenamefont {Hirschfeld},\ and\ \citenamefont {Scalapino}}]{graser2009near}%
  \BibitemOpen
  \bibfield  {author} {\bibinfo {author} {\bibfnamefont {S.}~\bibnamefont {Graser}}, \bibinfo {author} {\bibfnamefont {T.}~\bibnamefont {Maier}}, \bibinfo {author} {\bibfnamefont {P.}~\bibnamefont {Hirschfeld}},\ and\ \bibinfo {author} {\bibfnamefont {D.}~\bibnamefont {Scalapino}},\ }\bibfield  {title} {\bibinfo {title} {Near-degeneracy of several pairing channels in multiorbital models for the fe pnictides},\ }\href@noop {} {\bibfield  {journal} {\bibinfo  {journal} {New Journal of Physics}\ }\textbf {\bibinfo {volume} {11}},\ \bibinfo {pages} {025016} (\bibinfo {year} {2009})}\BibitemShut {NoStop}%
\bibitem [{\citenamefont {K\"onig}\ and\ \citenamefont {Coleman}(2019)}]{coleman2019coulomb}%
  \BibitemOpen
  \bibfield  {author} {\bibinfo {author} {\bibfnamefont {E.~J.}\ \bibnamefont {K\"onig}}\ and\ \bibinfo {author} {\bibfnamefont {P.}~\bibnamefont {Coleman}},\ }\bibfield  {title} {\bibinfo {title} {Coulomb problem in iron-based superconductors},\ }\href {https://doi.org/10.1103/PhysRevB.99.144522} {\bibfield  {journal} {\bibinfo  {journal} {Phys. Rev. B}\ }\textbf {\bibinfo {volume} {99}},\ \bibinfo {pages} {144522} (\bibinfo {year} {2019})}\BibitemShut {NoStop}%
\bibitem [{\citenamefont {Yang}\ \emph {et~al.}(2023{\natexlab{d}})\citenamefont {Yang}, \citenamefont {Sun}, \citenamefont {Hu}, \citenamefont {Xie}, \citenamefont {Miao}, \citenamefont {Luo}, \citenamefont {Chen}, \citenamefont {Liang}, \citenamefont {Zhu}, \citenamefont {Qu} \emph {et~al.}}]{yang2023orbital}%
  \BibitemOpen
  \bibfield  {author} {\bibinfo {author} {\bibfnamefont {J.}~\bibnamefont {Yang}}, \bibinfo {author} {\bibfnamefont {H.}~\bibnamefont {Sun}}, \bibinfo {author} {\bibfnamefont {X.}~\bibnamefont {Hu}}, \bibinfo {author} {\bibfnamefont {Y.}~\bibnamefont {Xie}}, \bibinfo {author} {\bibfnamefont {T.}~\bibnamefont {Miao}}, \bibinfo {author} {\bibfnamefont {H.}~\bibnamefont {Luo}}, \bibinfo {author} {\bibfnamefont {H.}~\bibnamefont {Chen}}, \bibinfo {author} {\bibfnamefont {B.}~\bibnamefont {Liang}}, \bibinfo {author} {\bibfnamefont {W.}~\bibnamefont {Zhu}}, \bibinfo {author} {\bibfnamefont {G.}~\bibnamefont {Qu}}, \emph {et~al.},\ }\bibfield  {title} {\bibinfo {title} {Orbital-dependent electron correlation in double-layer nickelate {La$_{3}$Ni$_{2}$O$_7$}},\ }\href@noop {} {\bibfield  {journal} {\bibinfo  {journal} {arXiv preprint arXiv:2309.01148}\ } (\bibinfo {year} {2023}{\natexlab{d}})}\BibitemShut {NoStop}%
\bibitem [{\citenamefont {Chen}\ \emph {et~al.}(2024)\citenamefont {Chen}, \citenamefont {Choi}, \citenamefont {Jiang}, \citenamefont {Mei}, \citenamefont {Jiang}, \citenamefont {Li}, \citenamefont {Agrestini}, \citenamefont {Garcia-Fernandez}, \citenamefont {Huang}, \citenamefont {Sun}, \citenamefont {Shen}, \citenamefont {Wang}, \citenamefont {Hu}, \citenamefont {Lu}, \citenamefont {Zhou},\ and\ \citenamefont {Feng}}]{chen2024electronic}%
  \BibitemOpen
  \bibfield  {author} {\bibinfo {author} {\bibfnamefont {X.}~\bibnamefont {Chen}}, \bibinfo {author} {\bibfnamefont {J.}~\bibnamefont {Choi}}, \bibinfo {author} {\bibfnamefont {Z.}~\bibnamefont {Jiang}}, \bibinfo {author} {\bibfnamefont {J.}~\bibnamefont {Mei}}, \bibinfo {author} {\bibfnamefont {K.}~\bibnamefont {Jiang}}, \bibinfo {author} {\bibfnamefont {J.}~\bibnamefont {Li}}, \bibinfo {author} {\bibfnamefont {S.}~\bibnamefont {Agrestini}}, \bibinfo {author} {\bibfnamefont {M.}~\bibnamefont {Garcia-Fernandez}}, \bibinfo {author} {\bibfnamefont {X.}~\bibnamefont {Huang}}, \bibinfo {author} {\bibfnamefont {H.}~\bibnamefont {Sun}}, \bibinfo {author} {\bibfnamefont {D.}~\bibnamefont {Shen}}, \bibinfo {author} {\bibfnamefont {M.}~\bibnamefont {Wang}}, \bibinfo {author} {\bibfnamefont {J.}~\bibnamefont {Hu}}, \bibinfo {author} {\bibfnamefont {Y.}~\bibnamefont {Lu}}, \bibinfo {author} {\bibfnamefont {K.-J.}\ \bibnamefont {Zhou}},\ and\ \bibinfo {author} {\bibfnamefont {D.}~\bibnamefont {Feng}},\ }\href@noop {}
  {\bibinfo {title} {Electronic and magnetic excitations in {La$_3$Ni$_2$O$_7$}}} (\bibinfo {year} {2024}),\ \Eprint {https://arxiv.org/abs/2401.12657} {arXiv:2401.12657 [cond-mat.supr-con]} \BibitemShut {NoStop}%
\end{thebibliography}%
\end{document}

\pagenumbering{Alph}

\title{Supplemental Material: Theory of magnetic excitations in multilayer nickelate superconductor La\textsubscript{3}Ni\textsubscript{2}O\textsubscript{7}}
\author{Steffen B\"otzel}
\affiliation{Institut f\"ur Theoretische Physik III, Ruhr-Universit\"at Bochum,
  D-44780 Bochum, Germany}
\author{Frank Lechermann}
\affiliation{Institut f\"ur Theoretische Physik III, Ruhr-Universit\"at Bochum,
  D-44780 Bochum, Germany}
\author{Jannik Gondolf}  
 \affiliation{Institut f\"ur Theoretische Physik III, Ruhr-Universit\"at Bochum,
  D-44780 Bochum, Germany} 
\author{Ilya M. Eremin}
\affiliation{Institut f\"ur Theoretische Physik III, Ruhr-Universit\"at Bochum,
  D-44780 Bochum, Germany}

%\linenumbers  %adds lines to reference changes
\maketitle
\subsection{Details on calculation of the susceptibility in the normal state}
The non-interacting multiorbital susceptibility in the superconducting state given by normal and anomalous Green's function
\begin{align}
    (\chi_0)_{\eta_1\eta_4}^{\eta_2\eta_3}(q) = \frac{T}{N} \sum_{k} 
[ F_{\eta_1\eta_3}(k+q) \bar{F}_{\eta_2\eta_4}(k) - G_{\eta_1\eta_2}(k+q) G_{\eta_3\eta_4}(k) ],
\label{Eq:nonintSus}
\end{align}
where we use the four notation $k=(\mathbf{k},k_z,i\omega_n)$ and shorthand indices $\eta = (l,o,s)$. In the normal state the anomalous Green's functions are zero. Here the non-interacting Green's function are given by
\begin{align}
    G_{\eta_1\eta_2}(\mathbf{k},k_z,i\omega_n) = \sum_{\mu}    \frac{a_{\eta_1,\mu}(\mathbf{k},k_z) a^*_{\eta_2,\mu}(\mathbf{k},k_z)}{i\omega_n - \epsilon_\mu(\mathbf{k})},
\end{align}
where $a_{\eta_1,\mu}(\mathbf{k},k_z)$ is a matrix element of the unitary transformation which diagonalizes the non-interacting part of the Hamiltonian connecting the combined sublattice-orbial-spin index $\eta$ with the $\mu$-th band $\epsilon_\mu(\mathbf{k})$. Since the phase factors are absorbed in the transformation, there is no $k_z$ dependence for the Green's function in band space. In the paramagnetic state the Green's functions are further diagonal in the spin index. Inserting this expression in Eq.~\ref{Eq:nonintSus}, the summation over the Matsubara frequencies can be done analytically by exploiting the residue theorem to find
\begin{align}
	(\chi_0)_{\eta_1\eta_4}^{\eta_2\eta_3}(\mathbf{q},q_z,i\nu_n)
	=  \sum_{\mathbf{k},k_z} \sum_{\mu\nu} 
 \frac{ f(\epsilon_{\mu}(\mathbf{k}))-f(\epsilon_{\nu}(\mathbf{\tilde{k}}))} 
 {\epsilon_{\nu}(\mathbf{\tilde{k}})-\epsilon_{\mu}(\mathbf{k})-i\nu_n}
  a_{\eta_1,\mu}(\mathbf{\tilde{k}},\tilde{k_z}) a^*_{\eta_2,\mu}(\mathbf{\tilde{k}},\tilde{k_z})a_{\eta_3,\nu}(\mathbf{k},k_z) a^*_{\eta_4,\nu}(\mathbf{k},k_z)
	\label{Eq:LindhardExp},
\end{align} 
using $\tilde{k} = k+q $ and $f$ denotes the Fermi-Dirac distribution function \cite{graser2009near}. The diagonalization of the initial Hamiltonian can be done in two steps by first block-diagonalizing to bonding and antibonding subspaces and then diagonalizing in both subspaces separately. Importantly, while there are four independent orbital indices, for the layer indices we can use $l_1 = l_4$ and $l_2 = l_3$. This is obvious from the fact that for the physical spin susceptibility we have to contract orbital and sublattice degrees of freedom at each of the free vertices. Further, the considered on-site interaction must also be diagonal in the sublattice degrees of freedom such that diagrams with two different layers appearing at the vertices are not contributing to the physical RPA susceptibility. For $l_1 = l_4$ the $k_z$ part of the phase factors included in $a_{\eta_1,\mu}(\mathbf{k}+\mathbf{q},k_z+q_z)$ and $a^*_{\eta_4,\nu}(\mathbf{k},k_z)$ must cancel and only the $q_z$ dependence remain. The same is true for the remaining matrix elements with $l_2 = l_3$ and therefore the susceptibility is of the form 
\begin{align}
    \hat{\chi}_0(\mathbf{q},q_z,i\nu_n) &=
    \begin{pmatrix}
		(\hat{\chi}_0)_{\parallel}(\mathbf{q},i\nu_n)	
        & (\hat{\chi}_0)_{\perp}(\mathbf{q},i\nu_n)	e^{iq_zd}  \\
		(\hat{\chi}_0)_{\perp}(\mathbf{q},i\nu_n)	 e^{-iq_zd}
        & (\hat{\chi}_0)_{\parallel}(\mathbf{q},i\nu_n)	
    \end{pmatrix},
    \label{eq:BilayerSus} 
\end{align}
as stated in the main text. The RPA susceptibility is then given by $\hat{\chi} = \hat{\chi_0} + \hat{\chi_0}\hat{U}\hat{\chi} + \hat{\chi_0}\hat{U}\hat{\chi}\hat{U}\hat{\chi} + ...$. Here it is important, that the interaction matrix must be block-diagonal as it only contains on-site interactions. From this expression, it follows that each appearing term and thus the full RPA susceptibility has the same form as Eq.~\ref{eq:BilayerSus}. It also follows that the RPA susceptibility can be decomposed in even and odd channels by applying the same transformation, which we used to transform from sublattice to bonding-antibonding space as displayed in Eq.~6 of the main text with the difference that here the susceptibility is a $2^2\times2^2$ matrix in orbital space as four different orbital indices appear in Eq.~\ref{Eq:LindhardExp}.

The interaction matrix is built using the on-site interaction part of the Hamiltonian, which includes intraorbital (interorbital) Hubbard repulsion $U$ ($U'$), Hund's coupling $J_{\rm H}$ and pair-hopping $J'$:
\begin{align}
	H^{(i)}_{int} = 
 \sum_{i,l} \left[
	U \sum_{o} 
	n_{ilo\uparrow}n_{ilo\downarrow}
	+ \sum_{o\neq o',s,s'} \left(
 \frac{U'}{2} 
	n_{ilos}n_{ilo's'}
  +
	 \frac{J_{\rm H}}{2}  
	c^{\dagger}_{ilos}c_{ilo's}c^{\dagger}_{ilo's'}
	c_{ilos'}
 \right)
	+ \frac{J'}{2} \sum_{o\neq o'} 
	c^{\dagger}_{ilo\uparrow}c_{io'\uparrow}c^{\dagger}_{ilo\downarrow}
	c_{ilo'\downarrow}
 \right]
\end{align}	
and the matrix blocks for each layer can be examined to be
\begin{align}
U_{o_2o_3}^{o_1o_4}  = 
	\begin{cases}
		{U}  &o_1=o_2=o_3=o_4  \\
		{U}' &o_1=o_2\neq o_3=o_4  \\ 
		{J}_{\rm H}  &o_1=o_4\neq o_2=o_3  \\
		{J}' &o_1=o_3\neq o_2=o_4 \\
		0  & otherwise.  
	\end{cases}
\end{align}

\begin{figure*}[t]
      \includegraphics[width=\linewidth]{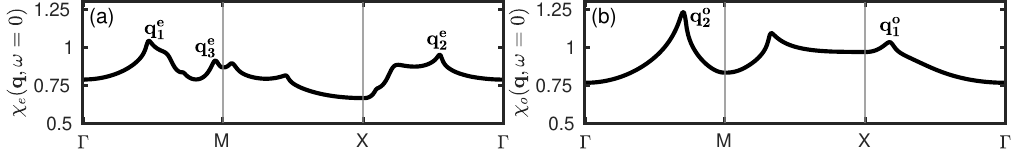}
	\caption{
  Calculated even (a) and odd (b) part of the static non-interacting spin susceptibilityat 80 K for the model from Ref.~\cite{LuoModel}.} \label{figS1}
\end{figure*}

For the physical susceptibility an analytic continuation $i\nu_n \rightarrow \omega + i\delta$ has to be performed. For the static susceptibilities the calculation is performed for $\delta \rightarrow 0$. The even and the odd channels of the RPA susceptibilities for the model from Ref.~\cite{LuoModel} are shown in Fig.1(c) and Fig.1(d), respectively, of the main text. For comparison we show the corresponding non-interacting susceptibilities in Fig.~\ref{figS1}(a) and Fig.~\ref{figS1}(b). 
%%%%%%%%%%%%%%%%%%%%%%%%%%%%%%%%%%%%%%%%%%%%%%%%%%%%%%%%%%%%%
%%%%%%%%%%%%%%%%%%%%%%%%%%%%%%%%%%%%%%%%%%%%%%%%%%%%%%%%%%%%%%

\subsection{Details on calculation of the susceptibility in the superconducting state}
For the superconducting state we add gap functions in the orbital or band space. The gap functions were taken from our previous calculationsRef.~\cite{lechermann2023electronic}. In the orbital space we start with a Bogoliubov-de-Gennes Hamiltonian
\begin{equation}
    H_{BdG}(\mathbf{k},k_z) =   
    \begin{pmatrix}
		\hat{H}(\mathbf{k},k_z)	
        & \hat{\Delta}(\mathbf{k},k_z)	 \\
		  \hat{\Delta}^\dagger(\mathbf{k},k_z)	 
        & -\hat{H}(\mathbf{k},k_z)	
    \end{pmatrix},
    \ \ \ \
        \hat{\Delta}(\mathbf{k},k_z) =
    \begin{pmatrix}
		\hat{\Delta}_{\parallel}(\mathbf{k})	
        & \hat{\Delta}_{\perp}(\mathbf{k})	e^{ik_zd}  \\
		\hat{\Delta}_{\perp}^\dagger(\mathbf{k}) e^{-ik_zd}
        & \hat{\Delta}(\mathbf{k})_{\parallel}
    \end{pmatrix}
    \label{eq:BilayerSus}
\end{equation}
for the basis $\Psi_\mathbf{k} = (\Psi_{\mathbf{k}\uparrow},\Psi_{-\mathbf{k}\downarrow})$ with $\Psi_{\mathbf{k}s} = (c_{\mathbf{k}Az^2s},c_{\mathbf{k}Ax^2-y^2s},c_{\mathbf{k}Bz^2s},c_{\mathbf{k}Bzx^2-y^2s})$ with $A$ and $B$ denoting the upper and lower layer as visualized in Fig.1(a) of the main text. As before, the Hamiltonian can be decomposed in bonding and antibonding parts and all the arguments concerning the bilayer structure remain valid. The normal and anomalous Green's functions are now calculated as 
\begin{align}
    G_{l_1o_1,l_2o_2}(\mathbf{k},k_z,i\omega_n) &= \sum_{\mu}    \frac{a_{l_1o_1\uparrow,\mu}(\mathbf{k},k_z) a^*_{l_2o_2\uparrow,\mu}(\mathbf{k},k_z)}{i\omega_n - E_\mu(\mathbf{k})}, \nonumber \\ F_{l_1o_1,l_2o_2}(\mathbf{k},k_z,i\omega_n) &= \sum_{\mu}    \frac{a_{l_1o_1\uparrow,\mu}(\mathbf{k},k_z) a^*_{l_2o_2\downarrow,\mu}(-\mathbf{k},k_z)}{i\omega_n - E_\mu(\mathbf{k})}
\end{align}
and inserted in Eq.~\ref{Eq:nonintSus}. Here $E_\mu(\mathbf{k})$ denotes the $\mu$-th band of the Bogoliubov-de-Gennes Hamiltonian. All remaining steps are similar to the normal state. 

For the imaginary part of dynamical susceptibilities we chose the damping factor $\delta = 1.5$ meV, i.e. one order of magnitude smaller than the characteristic size of the superconducting gap function $\Delta_0 = 15$ meV. This is an appropriate size to ensure that the spin resonances are still relatively sharp as the superconducting gap is the characteristic scale. A small $\delta$ requires a fine momentum resolution to avoid numerical inaccuracy. In particular, we chose a $2600\times2600$ momentum grid when the gap functions are included in band space ($2048\times2048$ for susceptiblity along $X-\Gamma-M$ high symmetry line shown in Fig.~\ref{figS2}) and a $1900\times1900$ grid when including the gap function in the orbital space. When introducing the gap function in band space, the transformation entering in Eq.~\ref{Eq:nonintSus} can be split further and can partially be done analytically, which allows to work with smaller matrices.

\subsection{Comparison of spin resonance for the $d_{xy}$ and $d_{x^2-y^2}$ superconducting gap functions}
\begin{figure*}[t]
      \includegraphics[width=\linewidth]{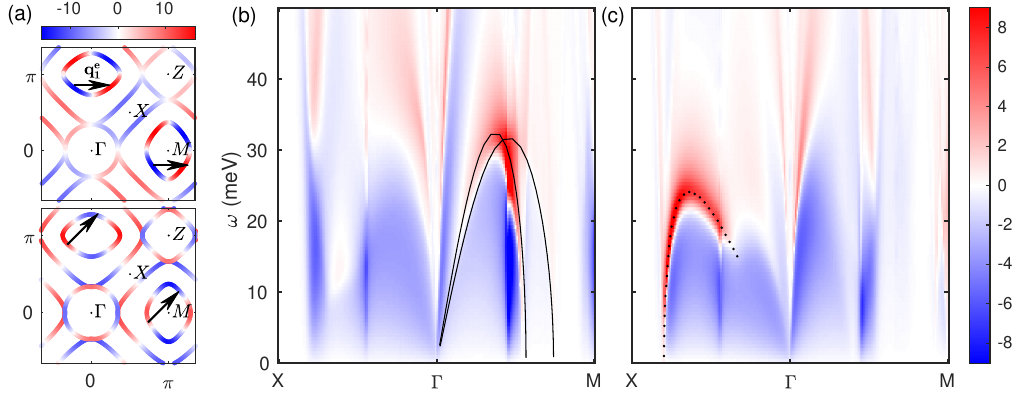}
	\caption{
  (a) The $d_{x^2-y^2}$ and $d_{xy}$ superconducting gap functions projected onto the normal state Fermi surface of Ref.~\cite{lechermann2023electronic}. (b) The difference of the even spin susceptibility between the normal state and superconducting state with the $d_{x^2-y^2}$ gap function calculated for $U=0.95 U_{\text{mag}}$ and $J = U/4$ at 80 K and 0 K, respectively, along the $X-\Gamma-M$ symmetry line. The black lines show the critical frequency branches arising due to the $\gamma$ pockets in the upper panel of (a). (c) Same as (b) but for the $d_{xy}$ gap function and parameters $U=0.95 U_{\text{mag}}$ and $J = U/7$.} \label{figS2}
\end{figure*}

Below we explore the differences of the $d_{x^2-y^2}$ and the $d_{xy}$ superconducting gaps, which are presented in Fig.\ref{figS2}(a) in the upper and lower panels, respectively. Here the gaps have been projected on the Fermi surface for the model from Ref.\cite{lechermann2023electronic}. Note that it is shown rotated with respect to the Fermi surface shown in the main text. In fact, for this model we do not restrict to hoppings within the bilayer sandwiches but allow all possible hoppings. However, we only consider a cut at $k_z = 0$ as explained in Ref.\cite{lechermann2023electronic}. Due to the structure of the Fmmm space group, this cut contains a $k_z = 0$ cut through the first BZ and a $k_z = \pi$ cut through the second BZ and so on \cite{Setyawan_2010}. The splitting in bonding and antibonding bands remains valid for this cut. With this procedure we include not only in-plane scattering but also scatterings between $k_z = 0$ and $k_z = \pi$ going beyond a two dimensional description of the band structure. 

In particular, the neighbouring $\alpha$ pockets and the neighbouring $\gamma$ pocket can be seen to be non-equivalent, which gives rise to two branches of critical frequencies. The critical frequency branch of a single pocket is given by $\Omega_c(\mathbf{q}) = |\Delta_{\mathbf{k}}| + |\Delta_{\mathbf{k}+\mathbf{q}}|$ with $\mathbf{k}$ and $\mathbf{k}+\mathbf{q}$ both being on the pocket and obeying a different sign for the superconducting gap. It has been used earlier to characterize dispersive spin resonances, which can only occur below the lowest critcal frequency for a particular momentum \cite{Eremin2005Novel}. This concept can be also used to distinguish the possible $d_{x^2-y^2}$ and $d_{xy}$ spin resonances. 

For the $d_{x^2-y^2}$ gap symmetry we have calculated the critical frequency branches for both $\gamma$ pockets and show it on top of the difference of calculated even spin susceptibility of the superconducting and normal state in Fig.~\ref{figS2}(b). Note that the dispersive peak nicely follows the lower of the branches and peaks near $\mathbf{q^e_1}$. The branches can be visualized by moving the shown arrows in the upper panel of Fig.\ref{figS2}(a) up and down while varying its length as the end and the start of the arrows always lay at the $\gamma$ pockets. The narrower branch belongs to the $\gamma$ pocket around $(\pi,0)$ in Fig.~\ref{figS2}(a). The same is shown for the $d_{xy}$ case in  Fig.~\ref{figS2}(c). For this case a branch along the $\Gamma-X$ symmetry line connects regions of different signs within the $\gamma$ pockets. This branch captures the dispersion of the peak. Here we show only the lower of the two almost identical branches as they are not crossing each other as for the $d_{x^2-y^2}$ case. The dominant role of the $\gamma$ band for the spin susceptibility is due  to its flatness.

The size of the gap functions is chosen such that its projection fits on what was found using a linearized gap function in Ref.~\cite{lechermann2023electronic} and that the magnitude of the projected gap function is about $\Delta_0 = 15$~meV. In particular, we chose $\Delta_\gamma = -1.5 \Delta_0 g_{x^2-y^2}$, $\Delta_\beta = 0.35 \Delta_0 g_{x^2-y^2}$ and $\Delta_\alpha = 0.4 \Delta_0 g_{x^2-y^2}$ with $g_{x^2-y^2}(\mathbf{k}) = \cos(k_x) - \cos(k_y)$ for the bands present at the Fermi surface for the $d_{x^2-y^2}$ symmetry. For the $d_{xy}$ symmetry, we use $\Delta_\gamma = -0.6 \Delta_0 g_{xy}$, $\Delta_\beta = -0.5 \Delta_0 g_{xy}$ and $\Delta_\alpha = 0.4 \Delta_0 g_{xy}$ with $g_{xy}(\mathbf{k}) = \cos(k_x+k_y)-\cos(k_x-k_y)$. Here $k_x$ and $k_y$ are chosen as displayed in Fig.~1 and Fig.~3 of the main text or in Fig.~\ref{figS3} of the supplementary.
%%%%%%%%%%%%%%%%%%%%%%%%%%%%%%%%%%%%%%%%%%%%%%%%%%%%%%%%%%%%%
%%%%%%%%%%%%%%%%%%%%%%%%%%%%%%%%%%%%%%%%%%%%%%%%%%%%%%%%%%%%%%

\subsection{Further details on the form of $s_\pm$-wave gap function and its feedback on spin resonance}
\begin{figure*}[t]
      \includegraphics[width=\linewidth]{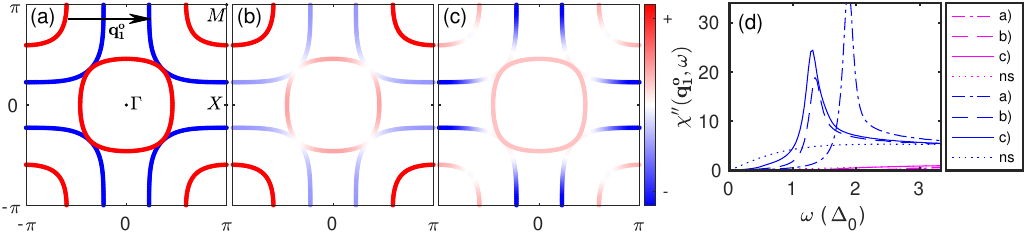}
	\caption{
  (a)-(c) Different gap functions realizing the $s$-wave symmetry projected onto the Fermi surface for the model~\cite{LuoModel}. (a) Only $\Delta^\perp_{x^2-y^2}$=$\Delta^\perp_{z^2}=\Delta_0$ are non-zero. (b) Only $\Delta^\perp_{z^2}=\Delta_0$ is non-zero. (c) Gap function introduced in the band space according to the solution found in Ref.~\cite{lechermann2023electronic} for the model given in Ref.~\cite{LuoModel}. (d) Similar to Fig.2a) of the main text with a comparison of the different gap functions used in (a),(b) and (c). Magenta and lines display the even and odd channel, respectively.} \label{figS3}
\end{figure*}

Since all the suggested version of the $s_\pm$ gap function in literature are similar but might differ a bit in certain details, we investigated how important these details are for our conclusion. In Fig.~\ref{figS3}(a)-(c) we are showing three different gap functions. In Fig.~\ref{figS3}(a) we show the case of $\Delta^\perp_{x^2-y^2}$=$\Delta^\perp_{z^2}=\Delta_0= 15$ meV for which the bonding bands bands have a constant gap of $\Delta_0$ and the antibonding $\beta$ band has a constant gap of $-\Delta_0$. In Fig.~\ref{figS3}(b) we only take $\Delta^\perp_{z^2}= \Delta_0$ to be non-zero. Consequently, the gap for the $\gamma$ band with $d_z^2$ character is quasi constant with the magnitude of 15 meV, whereas the $\alpha$ and the $\beta$ bands have smaller magnitudes, which resembles their mixed orbital character. In Fig.~\ref{figS3}(c) we introduced gaps in the band space in order to fit the gap functions found in Ref.~\cite{lechermann2023electronic} with highest magnitude at the $\beta$ band. In particular, we used an extended $s_\pm$ wave $\Delta_\gamma = -0.96 \Delta_0 + 0.72 \Delta_0 g_{x^2+y^2}$, $\Delta_\beta = 0.84 \Delta_0 - 0.96 \Delta_0 g_{x^2+y^2}$ and  $\Delta_\alpha = 0.48 \Delta_0$ with $g_{x^2+y^2} = \cos^2(k_x) + \cos^2(k_y)$. Due to sizeable orbital hybridization, this corresponds to a more complicated gap with interorbital and intralayer components in the orbital-sublattice space. In the main text we show the results for the most realistic third case. However, the conclusions are valid for all the considered cases. A strong and sharp spin resonance is appearing exclusively at $q_1^o$ in the odd channel of the spin susceptibility in the superconducting state. We compare the three cases for that scattering in Fig.~\ref{figS3}(d). The cases (b) and (c) appear at similar energy transfers with the latter having a bit larger intensity. For the (a) case, the spin resonance appears right below 2$\Delta_0 = 30$ meV as the gap magnitude for both gaps is $\pm\Delta_0$ and is of larger intensity. However, the qualitative predicted feature, i.e. a single large spin resonance at $q_1^o$ for the $_\pm$-wave symmetry of the superconducting gap is still valid. 

%%%%%%%%%%%%%%%%%%%%%%%%%%%%%%%%%%%%%%%%%%%%%%%%%%%%%%%%%%%%%
%%%%%%%%%%%%%%%%%%%%%%%%%%%%%%%%%%%%%%%%%%%%%%%%%%%%%%%%%%%%%%

\bibliography{literatur_supplementary}